\documentclass[%
  reprint,
  amsfonts,amssymb,amsmath,
  aps,
  pra
]{revtex4-1}

\usepackage[ascii]{inputenc}
\usepackage[T1]{fontenc}
\usepackage[english]{babel}
\usepackage{graphicx}
\usepackage{microtype}
\usepackage{bm}
\usepackage[pdftex,colorlinks]{hyperref}
\hypersetup{%
  linkcolor=blue,%
  citecolor=blue,%
  urlcolor=blue,%
  pdftitle={Stationary states in the many-particle description of Bose-Einstein
            condensates with balanced gain and loss},%
  pdfauthor={Dennis Dast, Daniel Haag, Holger Cartarius, J\"org Main,
             G\"unter Wunner}%
}

\DeclareMathOperator{\acos}{arccos}
\DeclareMathOperator{\atan}{arctan}
\newcommand{\PT}{\mathcal{PT}}
\newcommand{\ha}{\hat{a}}
\newcommand{\hH}{\hat{H}}
\newcommand{\hL}{\hat{L}}
\newcommand{\hA}{\hat{A}}
\newcommand{\hn}{\hat{n}}
\newcommand{\hrho}{\hat{\rho}}
\newcommand{\br}{\bm{r}}
\newcommand{\mrm}{\mathrm}
\newcommand{\diff}{\mathrm{d}}
\newcommand{\eul}{\mathrm{e}}
\newcommand{\dt}[1][]{\frac{\diff {#1}}{\diff t}}
\newcommand{\bra}[1]{\langle{#1}|}
\newcommand{\ket}[1]{|{#1}\rangle}
\newcommand{\mean}[1]{\langle{#1}\rangle}
\newcommand{\loss}{\mathrm{loss}}
\newcommand{\gain}{\mathrm{gain}}

\begin{document}

\title{Stationary states in the many-particle description of Bose-Einstein
       condensates with balanced gain and loss}

\author{Dennis Dast}
\email[]{dennis.dast@itp1.uni-stuttgart.de}
\author{Daniel Haag}
\author{Holger Cartarius}
\author{J\"org Main}
\author{G\"unter Wunner}

\affiliation{Institut f\"ur Theoretische Physik 1,
             Universit\"at Stuttgart, 70550 Stuttgart, Germany}

\date{\today}

\begin{abstract}
  Bose-Einstein condensates with balanced gain and loss can support stationary
  states despite the exchange of particles with the environment.
  In the mean-field approximation this is described by the $\PT$-symmetric
  Gross-Pitaevskii equation with real eigenvalues.
  In this work we study the role of stationary states in the appropriate
  many-particle description.
  It is shown that without particle interaction there exist two non-oscillating
  trajectories which can be interpreted as the many-particle equivalent of the
  stationary $\PT$-symmetric mean-field states.
  Furthermore the system has a non-equilibrium steady state which acts as an
  attractor in the oscillating regime.
  This steady state is a pure condensate for strong gain and loss contributions
  if the interaction between the particles is sufficiently weak.
\end{abstract}

\maketitle

\section{Introduction}
\label{sec:introduction}

Non-Hermitian Hamiltonians can be used as an elegant effective description of
open quantum systems~\cite{Moiseyev11a}.
A special class of non-Hermitian Hamiltonians that gained much interest since
the seminal paper by Bender and Boettcher~\cite{Bender98a} are $\PT$-symmetric
Hamiltonians which fulfill $[\hH, \PT] = 0$ with the parity reflection operator
$\mathcal{P}$ and the time reversal operator $\mathcal{T}$.
The outstanding property of these Hamiltonians is that they can support an
entirely real eigenvalue spectrum despite being non-Hermitian \cite{Bender98a,
Bender02a}.

In position space non-Hermitian Hamiltonians can be achieved by introducing
complex potentials.
For Bose-Einstein condensates imaginary parts of the potential have a clear
physical interpretation as a localized gain and loss of
particles~\cite{Kagan98a}.
If the system is $\PT$ symmetric the complex potential must fulfill $V(\br) =
V^*(-\br)$.
This condition means that the gain and loss contributions are balanced in such
a way that stationary states with real eigenvalues can exist.

Due to its simplicity a double-well potential where particles are removed from
one well and injected into the other is an ideal system to study the
properties of $\PT$-symmetric Bose-Einstein condensates.
Both the controlled incoupling and outcoupling of particles have already been
demonstrated experimentally.
The localized particle loss can be realized by a focused electron beam which
ionizes the atoms whereupon they escape the trapping
potential~\cite{Gericke08a, Wurtz09a, Barontini13a}.
Incoupling of particles can be realized by feeding from a second
condensate~\cite{Robins08a} in a Raman superradiance-like
process~\cite{Doring09a, Schneble04a, Yoshikawa04a}.

This system has been extensively studied in the mean-field limit where it is
described by the $\PT$-symmetric Gross-Pitaevskii equation using discrete
matrix models~\cite{Graefe12a, Haag15a}, double-delta
potentials~\cite{Cartarius12b, Cartarius12a} and spatially extended double
wells~\cite{Dast13a, Dast13b, Haag14a, Dizdarevic15a}.
In these works it was shown that the system supports stationary $\PT$-symmetric
states, exhibits intriguing dynamical properties, and possesses a rich
eigenvalue structure, including various exceptional points.
Furthermore, proposals to simulate an effective $\PT$-symmetric double well by
embedding it into a larger Hermitian structure have been formulated in the
mean-field approximation~\cite{Kreibich13a, Kreibich14a}.

However, if the exchange of particles with the environment plays such a crucial
role, one has to expect effects that cannot be captured by the mean-field
approximation in which a pure condensate described by a product of identical
single-particle states is assumed~\cite{Ruostekoski98a, Syassen08a,
Witthaut08a, Witthaut09a, Witthaut11a, Barmettler11a, Labouvie16a}.
In fact it was shown that in case of a two-mode Bose-Einstein condensate with
balanced gain and loss the purity of the condensate oscillates, i.e., it
periodically drops to small values but then is almost completely
restored~\cite{Dast16a, Dast16b}.
These purity oscillations are experimentally accessible by measuring the
average contrast in an interference experiment~\cite{Dast16a}.
This shows that to fully understand Bose-Einstein condensates with balanced
gain and loss a description that goes beyond the mean field is required.

The most characteristic property of $\PT$-symmetric systems is the possibility
of the existence of an entirely real eigenvalue spectrum and, consequently, the
existence of stationary states despite the Hamiltonian being non-Hermitian.
Hence, it is natural to ask whether the many-particle system with balanced gain
and loss also supports stationary states and whether these stationary states
are connected to the $\PT$-symmetric stationary states of the mean-field limit.

In this work we study a two-mode Bose-Einstein condensate with balanced gain
and loss described by a master equation in Lindblad form which is presented in
Sec.~\ref{sec:system}.
It is shown that the stationary solutions of the mean-field limit are not the
stationary states of the master equation but instead they are located close to
non-oscillatory trajectories, whose properties are discussed in
Sec.~\ref{sec:stat_nonosci}.
In Sec.~\ref{sec:stat_ss} the steady state of the many-particle system, defined
as $\dt \hrho = 0$, is calculated and its influence on the dynamics of the
system is studied.
It is a non-equilibrium steady state since the system is subject to particle
gain and loss, and thus, a compensating particle current has to be present.

\section{Two-mode system with balanced gain and loss}
\label{sec:system}

The many-particle description of a Bose-Einstein condensate with balanced gain
and loss in a double-well potential was introduced in~\cite{Dast14a} using a
two-mode approximation, and its dynamical properties were discussed
in~\cite{Dast16a} and~\cite{Dast16b}.
In this section the properties required for this work are briefly recapped.

The master equation in Lindblad form~\cite{Breuer02a, Anglin97a} describing
this system reads
\begin{equation}
  \dt \hrho = -i [\hH,\hrho]
  + \gamma_\loss \mathcal{L}(\ha_1) \hrho
  + \gamma_\gain \mathcal{L}(\ha_2^\dagger) \hrho,
  \label{eq:balancedMasterEq}
\end{equation}
where the coherent dynamics in the double well is given by the Bose-Hubbard
Hamiltonian~\cite{Fisher89a, Jaksch98a} for two lattice sites
\begin{equation}
  \hH = -J \left( \ha_1^\dagger \ha_2 + \ha_2^\dagger \ha_1 \right)
  + \frac{U}{2} \left( \ha_1^\dagger \ha_1^\dagger \ha_1 \ha_1
  + \ha_2^\dagger \ha_2^\dagger \ha_2 \ha_2 \right),
  \label{eq:boseHubbardTwoMode}
\end{equation}
and the localized loss at site 1 and gain at site 2 are introduced by the
Lindblad terms
\begin{align}
  \mathcal{L}(\ha_1) \hrho &= -\frac{1}{2}
  \left(\ha_1^\dagger \ha_1 \hrho + \hrho \ha_1^\dagger \ha_1
  - 2 \ha_1 \hrho \ha_1^\dagger \right),
  \label{eq:lindbladLoss}\\
  \mathcal{L}(\ha_2^\dagger) \hrho &= -\frac{1}{2}
  \left(\ha_2 \ha_2^\dagger \hrho + \hrho \ha_2 \ha_2^\dagger
  - 2 \ha_2^\dagger \hrho \ha_2 \right).
  \label{eq:lindbladGain}
\end{align}
The parameter $J$ is the tunneling strength between the two sites.
The on-site interaction strength between the particles is given by $U$.
By demanding that the time derivative of the total particle number vanishes
when the expectation values of the particle number are identical at both sites,
we obtain a relation for the gain and loss rates $\gamma_\gain$ and
$\gamma_\loss$~\cite{Dast14a},
\begin{equation}
  \gamma_\loss = \frac{N_0+2}{N_0}\gamma_\gain \equiv \gamma,
  \label{eq:gainlossratio}
\end{equation}
with $N_0$ the initial number of atoms.
In the following $\gamma_\loss$ and $\gamma_\gain$ are always chosen such that
they fulfill this relation, and the abbreviation $\gamma$ will be used to
characterize the strength of the gain and loss.

As shown in~\cite{Dast16b} the Bogoliubov backreaction method \cite{Anglin01a,
Vardi01a} can describe the dynamics of this open quantum system accurately for
a limited time span.
The basic idea is to formulate equations of motion for the first-order moments
$\mean{\ha_j^\dagger \ha_k}$ which couple to the second-order moments
$\mean{\ha_j^\dagger \ha_k \ha_l^\dagger \ha_m}$ via the interaction term.
The coupling of the second-order moments to the third-order moments is
eliminated by the approximation
\begin{align}
  & \mean{\ha_i^\dagger \ha_j \ha_k^\dagger \ha_l
  \ha_m^\dagger \ha_n} \approx
  \mean{\ha_i^\dagger \ha_j \ha_k^\dagger
  \ha_l}\mean{\ha_m^\dagger \ha_n}
  \notag\\
  &\quad + \mean{\ha_i^\dagger \ha_j \ha_m^\dagger
  \ha_n}\mean{\ha_k^\dagger \ha_l}
  + \mean{\ha_k^\dagger \ha_l \ha_m^\dagger
  \ha_n}\mean{\ha_i^\dagger \ha_j}
  \notag\\
  &\quad - 2 \mean{\ha_i^\dagger \ha_j}
  \mean{\ha_k^\dagger \ha_l} \mean{\ha_m^\dagger \ha_n},
  \label{eq:thirdordermoments}
\end{align}
such that a closed set of equations of motion is obtained for the first- and
second-order moments.

For the two-mode system this method is most conveniently formulated using the
Bloch representation, where the expectation values $s_j = 2\mean{\hL_j}$ and $n
= \mean{\hn}$ of the four Hermitian operators
\begin{subequations}
  \begin{align}
    \hL_x &= \frac{1}{2} (\ha_1^\dagger \ha_2 + \ha_2^\dagger \ha_1),&
    \hL_y &= \frac{i}{2} (\ha_1^\dagger \ha_2 - \ha_2^\dagger \ha_1),\\
    \hL_z &= \frac{1}{2} (\ha_2^\dagger \ha_2 - \ha_1^\dagger \ha_1),&
    \hn &= \ha_1^\dagger \ha_1 + \ha_2^\dagger \ha_2,
  \end{align}
  \label{eq:blochoperators}%
\end{subequations}
and the covariances
\begin{equation}
  \Delta_{j k} = \mean{\hA_j \hA_k + \hA_k \hA_j} - 2 \mean{\hA_j}\mean{\hA_k}
  \label{eq:blochcov}
\end{equation}
with $\hA_{j} \in \{\hL_x, \hL_y, \hL_z, \hn\}$ are used.
The resulting set of equations of motions of the first-order moments reads
\begin{subequations}
  \begin{align}
    \dot{s}_x &= -U (s_y s_z + 2\Delta_{yz}) - \gamma_- s_x, \\
    \dot{s}_y &= 2 J s_z + U(s_x s_z + 2\Delta_{xz}) - \gamma_- s_y, \\
    \dot{s}_z &= -2 J s_y + \gamma_+ n - \gamma_- s_z + \gamma_\gain, \\
    \dot{n}   &= -\gamma_- n + \gamma_+ s_z + \gamma_\gain,
  \end{align}
  \label{eq:dspdmbloch}%
\end{subequations}
with $\gamma_- = (\gamma_\loss-\gamma_\gain)/2$ and $\gamma_+ =
(\gamma_\loss+\gamma_\gain)/2$.
The ten equations of motion of the covariances can be found in~\cite{Dast16b}.
Their specific form is not required for the understanding of this paper.
The Bloch representation is well suited to discuss the purity of a condensate
since it takes the particularly simple form
\begin{equation}
  P = \frac{s_x^2 + s_y^2 + s_z^2}{n^2}.
  \label{eq:purityTwomode}
\end{equation}
This quantity signals the existence of off-diagonal long-range
order~\cite{Penrose56a, Yang62a}, and thus measures how close the condensate
is to a pure condensate, described by a product of identical single-particle
states.

In the non-interacting limit the equations of motion for the first- and
second-order moments decouple and an analytic solution can be
obtained~\cite{Dast16b}.
Note that the results obtained in this limit are exact, however, they are not
equivalent to the $\PT$-symmetric Gross-Pitaevskii equation since
Eq.~\eqref{eq:dspdmbloch} can predict fragmentation of the condensate.
In the oscillatory regime $\gamma_+^2 < 4J^2$ the solution reads
\begin{subequations}
  \begin{align}
    s_x(t) &= \alpha_1 + \kappa_1 \eul^{-\gamma_- t},\\
    s_y(t) &= \alpha_2 + [\gamma_+ \kappa_2
              + 2 J \kappa_3 \cos(\omega t - \kappa_4)] \eul^{-\gamma_- t},\\
    s_z(t) &= \alpha_3 - \omega \kappa_3 \sin(\omega t - \kappa_4)
              \eul^{-\gamma_- t},\\
    n(t)   &= \alpha_4 + [2J \kappa_2
              + \gamma_+ \kappa_3 \cos(\omega t - \kappa_4)]
              \eul^{-\gamma_- t}
  \end{align}
  \label{eq:solosci}%
\end{subequations}
with $\omega = \sqrt{4J^2 - \gamma_+^2}$ and the constant inhomogeneous
solution, i.e.\ the steady state, being given by
\begin{equation}
  \bm{\alpha} =
  \frac{\gamma_+^2-\gamma_-^2}{4J^2-\gamma_+^2+\gamma_-^2}
  \begin{pmatrix}
    0 \\
    \frac{2J}{\gamma_-} \\
    1 \\
    1 + \frac{4J^2}{\gamma_-(\gamma_+ + \gamma_-)}
  \end{pmatrix}.
  \label{eq:partsol}%
\end{equation}
The parameters $\kappa_i$ in Eq.~\eqref{eq:solosci} are defined by the
initial state $(s_x(0),\ s_y(0),\ s_z(0),\ n(0))^T$,
\begin{subequations}
  \begin{align}
    \kappa_1 &= s_x(0),\\
    \kappa_2 &= \frac{2J (n(0) - \alpha_4) - \gamma_+ (s_y(0) - \alpha_2)}
    {\omega^2},\\
    \kappa_3 &= k_1 \sqrt{1+\left(\frac{k_2}{k_1}\right)^2},\\
    \kappa_4 &= \atan\left(\frac{k_2}{k_1}\right),\\
    \intertext{with the abbreviations}
    k_1 &= \frac{2J (s_y(0) - \alpha_2) - \gamma_+ (n(0) - \alpha_4)}
    {\omega^2},\\
    k_2 &= \frac{s_z(0) - \alpha_3}{\omega}.
  \end{align}
\end{subequations}

The mean-field approximation of the two-mode master equation with balanced gain
and loss~\eqref{eq:balancedMasterEq} is obtained in the limit $N_0 \to \infty$
but with constant macroscopic interaction strength $g = U (N_0 - 1) \approx U
N_0$.
To obtain the Gross-Pitaevskii equation an initially pure condensate is
assumed although mean-field theories exist where the bosons reside in several
different single-particle states~\cite{Cederbaum03a}.
In the Gross-Pitaevskii-like mean-field limit the condensate is described by a
product of identical single-particle states, thus, it is defined by two complex
mean-field coefficients $\psi = (c_1, c_2)^T$.
It was shown in~\cite{Dast14a} that the mean-field limit is the discrete
$\PT$-symmetric Gross-Pitaevskii equation~\cite{Graefe08b, Graefe12a}
\begin{subequations}
  \begin{align}
    i \dt c_1 &= - J c_2 + g |c_1|^2 c_1
      - i \frac{\gamma}{2} c_1,\\
    i \dt c_2 &= - J c_1 + g |c_2|^2 c_2
      + i \frac{\gamma}{2} c_2.
  \end{align}
  \label{eq:discreteGPE}%
\end{subequations}
After normalizing the state and choosing an arbitrary global phase only two
degrees of freedom remain for a mean-field state which can be expressed using
the angles $\varphi$ and $\vartheta$,
\begin{subequations}
  \begin{align}
    \varphi &= \arg(c_1 c_2^*),\\
    \vartheta &= \acos(1 - 2|c_1|^2),
  \end{align}
  \label{eq:phithetaTrans}%
\end{subequations}
chosen in such a way that the Bloch representation of the first-order moments
of a mean-field state with $N_0$ particles is given by the spherical
coordinates
\begin{subequations}
  \begin{align}
    s_x &= N_0 \sin(\vartheta) \cos(\varphi),\\
    s_y &= N_0 \sin(\vartheta) \sin(\varphi),\\
    s_z &= N_0 \cos(\vartheta).
  \end{align}
  \label{eq:purestateBloch}%
\end{subequations}
In this representation the two stationary $\PT$-symmetric states of
Eq.~\eqref{eq:discreteGPE} read
\begin{subequations}
  \begin{align}
    \varphi &= \frac{\pi}{2} \mp \acos\left(\frac{\gamma}{2J}\right),\\
    \vartheta &= \frac{\pi}{2},
  \end{align}
  \label{eq:stationaryStatesGpeAngles}%
\end{subequations}
where the upper (lower) sign is for the ground (excited) state of the system.

\section{Non-oscillatory states}
\label{sec:stat_nonosci}

In this section the non-oscillatory states of the many-particle dynamics are
discussed in the non-interacting limit $U = 0$ and it is shown that they can be
interpreted as the equivalent of the $\PT$-symmetric stationary states of the
Gross-Pitaevskii equation.

We start from the solutions in the oscillatory regime $\gamma_+^2 < 4J^2$ given
by Eq.~\eqref{eq:solosci}.
As can be directly seen all oscillatory terms vanish for $\kappa_3 = 0$ and a
non-oscillating trajectory is given by
\begin{subequations}
  \begin{align}
    s_x(t) &= \kappa_1 \eul^{-\gamma_- t},\\
    s_y(t) &= \alpha_2 + \gamma_+ \kappa_2 \eul^{-\gamma_- t},\\
    s_z(t) &= \alpha_3,\\
    n(t)   &= \alpha_4 + 2J \kappa_2 \eul^{-\gamma_- t},
  \end{align}
\end{subequations}
where the only remaining time dependence is the exponential decay towards the
steady state $\bm{\alpha}$.

The condition $\kappa_3=0$ can be fulfilled by pure initial states which are
expressed by the two angles $\varphi$ and
$\vartheta$ as defined in Eq.~\eqref{eq:purestateBloch}.
In the following it will be shown that  for a specific initial particle number
$N_0$ there are two pure states which fulfill this condition.
They are obtained by solving the set of equations
\begin{subequations}
  \begin{align}
    s_x(0) &= \kappa_1 = N_0 \sin\vartheta \cos\varphi,\\
    s_y(0) &= \alpha_2 + \gamma_+ \kappa_2 = N_0 \sin\vartheta \sin\varphi,\\
    s_z(0) &= \alpha_3 = N_0 \cos\vartheta,\\
    n(0)   &= \alpha_4 + 2J \kappa_2 = N_0.
    \label{eq:nonosciInitN}
  \end{align}
\end{subequations}
The angle $\vartheta$ is directly obtained from the third equation,
\begin{equation}
  \vartheta = \acos\left(\frac{\alpha_3}{N_0}\right)
  = \acos\left(\frac{\frac{1}{N_0+2} \gamma^2}
  {4J^2 - \frac{N_0}{N_0+2}\gamma^2}\right),
  \label{eq:nonOsciThetaLin}
\end{equation}
where the steady state~\eqref{eq:partsol} and the relation for balanced gain
and loss~\eqref{eq:gainlossratio} were used.

Before calculating the remaining angle $\varphi$, the two parameters $\kappa_1$
and $\kappa_2$ must be determined.
Equation~\eqref{eq:nonosciInitN} yields the value of the parameter $\kappa_2$,
\begin{equation}
  \kappa_2 = \frac{1}{2J}(N_0 - \alpha_4),
  \label{eq:nonosciKappa2}
\end{equation}
whereas two values for $\kappa_1$ follow from the requirement that the initial
state must be pure,
\begin{equation}
  \kappa_1 = s_x(0) = \pm \sqrt{N_0^2 - s_y(0)^2 - s_z(0)^2}.
  \label{eq:nonosciKappa1}
\end{equation}
Since a short calculation shows that $s_y(0) > 0$ in the parameter regime
considered, and the two values for $s_x(0)$ only differ in their sign, the
angle $\varphi$ takes the following two values,
\begin{align}
  &\varphi_\mp
  = \frac{\pi}{2} \mp
  \acos \Bigg[ \frac{\gamma}{2J}
    \left(4J^2 - \frac{(N_0+1)^2}{(N_0+2)^2} \gamma^2\right)
  \notag\\
  &\left(4J^2 - \frac{N_0+1}{N_0+2} \gamma^2\right)^{-1/2}
  \left(4J^2 - \frac{N_0-1}{N_0+2} \gamma^2\right)^{-1/2} \Bigg].
  \label{eq:nonOsciPhiLin}
\end{align}
\begin{figure}[t]
  \centering
  \includegraphics[width=0.75\columnwidth]{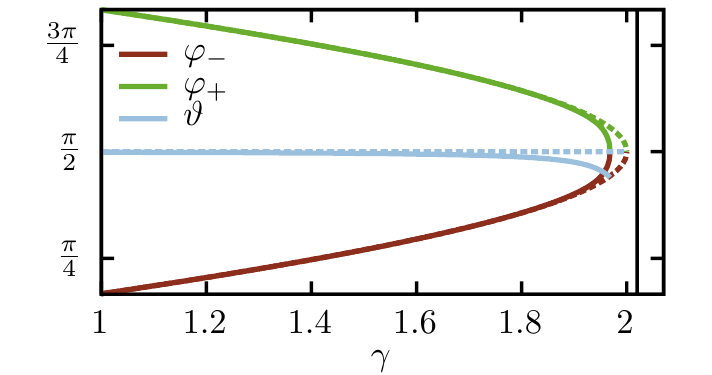}
  \caption{%
    The non-oscillatory solutions (solid lines) and the stationary
    $\PT$-symmetric solutions (dotted lines) as a function of the gain-loss
    parameter $\gamma$ for $N_0 = 100$ and $U = 0$.
    The non-oscillatory states coalesce and vanish slightly before the
    $\PT$-symmetric solutions.
    Both types of states vanish before the critical point $\gamma_+ = 2J$
    (i.e.\ $\gamma = 2J \frac{N_0+2}{N_0+1}$) indicated by the vertical black
    line, at which the oscillatory regime ends.
  }
  \label{fig:stationaryNonosciLin}
\end{figure}

To interpret the non-oscillatory states as the many-particle equivalent of the
stationary $\PT$-symmetric states, it is necessary that they become equal in
the limit $N_0 \to \infty$.
In this limit the argument of the arccosine function in
Eq.~\eqref{eq:nonOsciThetaLin} vanishes,
\begin{equation}
  \lim_{N_0 \to \infty} \vartheta = \frac{\pi}{2},
\end{equation}
and in Eq.~\eqref{eq:nonOsciPhiLin} only the first fraction remains,
\begin{equation}
  \lim_{N_0 \to \infty} \varphi_\mp =
  \frac{\pi}{2} \mp \acos\left(\frac{\gamma}{2J}\right).
  \label{eq:nonosciPhiMeanfield}
\end{equation}
This shows that for $N_0 \to \infty$ indeed the non-oscillatory states become
equal to the $\PT$-symmetric states given by
Eq.~\eqref{eq:stationaryStatesGpeAngles}.
To be more precise, the state $(\varphi_-, \vartheta)$ becomes the ground state
and $(\varphi_+, \vartheta)$ the excited state of the mean-field system.

Plotting the spherical coordinates as a function of $\gamma$ for a constant
particle number $N_0 = 100$ as done in Fig.~\ref{fig:stationaryNonosciLin}
yields the characteristic structure of an exceptional point of order $2$.
Note that in Fig.~\ref{fig:stationaryNonosciLin} as well as in the following
figures the value of $J$ is set to 1, without loss of generality.
The two non-oscillatory states (solid lines) coalesce slightly before the value
of $\gamma$ at which the $\PT$-symmetric ground and excited state (dotted
lines) become equal.

Note that both the non-oscillatory and the stationary $\PT$-symmetric states
coalesce at values of $\gamma$ that are slightly smaller than the critical
point $\gamma_+^2 = 4J^2$  at which the oscillatory regime ends, as indicated
by the vertical black line in Fig.~\ref{fig:stationaryNonosciLin}.
Only in the mean-field limit $N_0 \to \infty$, the coalescence of all states
and the critical point coincide.

\begin{figure}[t]
  \centering
  \includegraphics[width=\columnwidth]{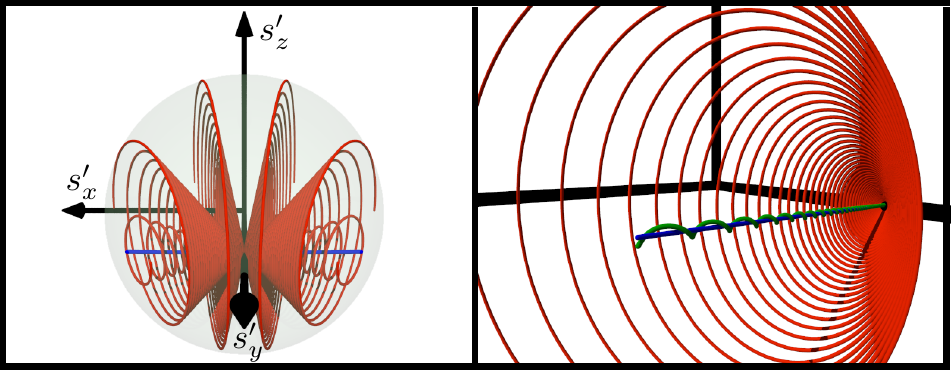}
  \caption{%
    Time evolution of the reduced quantities~\eqref{eq:reducedBlochVector} in
    the non-interacting limit $U = 0$.
    All trajectories encircle the two non-oscillatory states (blue lines),
    thus, motivating their interpretation as the many-particle equivalent of
    the $\PT$-symmetric solutions (green lines in the right panel).
    The right panel shows a single cone with $s_x' > 0$ in more detail.
    The initial particle number is $N_0 = 100$ and the gain-loss parameter is
    $\gamma = 1.5$.
  }
  \label{fig:blochNonosci}
\end{figure}

The special role of the non-oscillatory states becomes clearly evident in
Fig.~\ref{fig:blochNonosci}, where the trajectories of the reduced components,
\begin{equation}
  s_{x,y,z}'(t) = \frac{s_{x,y,z}(t)}{n(t)},
  \label{eq:reducedBlochVector}
\end{equation}
are shown.
In this representation the squared norm of the vector $(s_x',\
s_y',\ s_z')^T$ is equal to the purity of the state $P = s_x'^2 +
s_y'^2 + s_z'^2$.

The left panel shows six trajectories for $\gamma=1.5$ and $N_0 = 100$ in red
lines which start on the surface of the sphere (which means they are initially
pure) at $\vartheta = \pi/2$ and $\varphi$ equally distributed between $0$ and
$2\pi$.
Furthermore, the two non-oscillating trajectories are plotted as blue lines.
One immediately recognizes that the dynamics is symmetric with respect to the
$s_y'$-$s_z'$ plane.
This can be understood by replacing $\kappa_1$ with $-\kappa_1$ in the general
solution given by Eq.~\eqref{eq:solosci}.
Then one obtains exactly the same trajectory with opposite sign of the
component $s_x(t)$, which leads to this symmetry.

Every trajectory has a structure similar to a cone and finally reaches the
steady state $\bm{\alpha}$ given by Eq.~\eqref{eq:partsol}, which lies
approximately on the $s_y'$ axis ($s_y' \approx 0.75$, corresponding to $P
\approx 0.56$).
Within the cones lie the trajectories of the non-oscillatory states, which are
encircled by all oscillating trajectories.
Since $\gamma_-$ is the exponential decay rate towards the steady state, the
decay is faster for bigger values of $\gamma$ resulting in less narrow
windings.
If $\gamma$ is increased, the non-oscillatory states on the surface of the
sphere are moved towards the $s_y'$ axis, i.e., approach $\varphi = \pi/2$
(cf.\ Fig.~\ref{fig:stationaryNonosciLin}).

The right panel of Fig.~\ref{fig:blochNonosci} shows the inner cone and the
non-oscillatory state with $s_x' > 0$ of the left panel in more detail.
Furthermore the trajectory of the $\PT$-symmetric ground state is added (green
line).
This trajectory encircles the non-oscillatory state at a very small distance.

In the mean-field limit initially pure states will stay pure for all times.
Thus, they will not leave the surface of the sphere and the two $\PT$-symmetric
states are elliptic fixed points, which are encircled by all trajectories.
In the many-particle system such fixed points do not exist, instead a
non-oscillating trajectory exists which is encircled by all states.
In this sense, the non-oscillatory states can be interpreted as the
many-particle equivalent of the $\PT$-symmetric stationary states.

If interaction between the particles is taken into account, the $\PT$-symmetric
states still show only weakly pronounced oscillations.
However, no clearly distinguished non-oscillating trajectories exist in their
vicinity.
Therefore, it is not possible to define a many-particle equivalent of the
$\PT$-symmetric states as done in the non-interacting limit.

\section{Non-equilibrium steady state}
\label{sec:stat_ss}

In this section the non-equilibrium steady state of the two-mode system with
balanced gain and loss is studied.
It is defined as a constant solution of the master equation, $\dt \hrho = 0$.
Before investigating the steady state numerically in the presence of
interaction, the analytically solvable non-interacting limit is discussed.

\subsection{Non-interacting limit}
\label{sec:stat_ss_linear}

In the non-interacting limit a constant solution, i.e.\ a steady state, exists
for $\gamma_+^2 \neq 4J^2 + \gamma_-^2$ and is given by Eq.~\eqref{eq:partsol}.
Furthermore this solution is an attractor in the parameter regime $\gamma_+^2 <
4J^2 + \gamma_-^2$ that every trajectory will finally reach.
Due to the prefactor the components $\alpha_2$, $\alpha_3$, and $\alpha_4$
diverge for $\gamma_+^2 \to 4J^2 + \gamma_-^2$.

The purity of the steady state can be calculated using
Eq.~\eqref{eq:purityTwomode},
\begin{equation}
  P = \frac{\alpha_1^2 + \alpha_2^2 + \alpha_3^2}{\alpha_4^2}
  = \frac{4J^2 + \gamma_-^2}{\left(\gamma_- + \frac{4J^2}{\gamma_+ +
  \gamma_-}\right)^2}.
  \label{eq:ss_purity}
\end{equation}
For every physical state the purity must not exceed one, therefore the steady
state is only a physical state in the parameter regime in which $P \leq 1$
holds, which is fulfilled for $\gamma_+^2 \leq 4J^2 + \gamma_-^2$.
This result is consistent with the dynamical behavior:
The steady state has a purity smaller than one, thus being physical, in the
parameter regime $\gamma_+^2 < 4J^2 + \gamma_-^2$, in which it acts as an
attractor.
At the critical point $\gamma_+^2 = 4J^2 + \gamma_-^2$ no constant solution
exists.
Although for $\gamma_+^2 > 4J^2 + \gamma_-^2$ the state $\bm{\alpha}$ is a
constant solution of the equations of motion, it is no longer physical since
its purity exceeds one.
Such a state cannot be reached dynamically which is consistent with the fact
that $\bm{\alpha}$ is no longer an attractor.

Using the relation for balanced gain and loss~\eqref{eq:gainlossratio}, the
purity of the steady state can be reformulated,
\begin{equation}
  P = \frac{\gamma^2}{4J^2} \frac{(N_0 + 2)^2 +
  \frac{\gamma^2}{4J^2}}{\left[(N_0 + 2) +
  \frac{\gamma^2}{4J^2}\right]^2}.
  \label{eq:ss_purity_exact}
\end{equation}
Since $N_0 \gg \gamma^2/4J^2$ the purity increases approximately
quadratically with $\gamma$
\begin{equation}
  P \approx \frac{\gamma^2}{4J^2},
  \label{eq:ss_purity_approx}
\end{equation}
and for $\gamma = 2J$ the steady state is a perfectly pure condensate
with $P=1$.

Due to $\gamma_- = \gamma/(N_0+2)$ the components $\alpha_2$ and
$\alpha_4$ are much larger than the component $\alpha_3$.
This can be clearly seen by looking at the components divided by the particle
number of the steady state $\alpha_4$,
\begin{subequations}
  \begin{align}
    \frac{\alpha_2}{\alpha_4} &= \frac{\gamma}{2J}
    \frac{N_0+2}{(N_0+2) + \frac{\gamma^2}{4J^2}},\\
    \frac{\alpha_3}{\alpha_4} &= \frac{\gamma}{2J}
    \frac{\frac{\gamma}{2J}}{(N_0+2) + \frac{\gamma^2}{4J^2}}.
  \end{align}
\end{subequations}
Consequently the purity is almost exclusively determined by $\alpha_2$.
This component is proportional to the tunneling current from site 2 to site 1,
$j_{2\to1} = J \alpha_2$.
Thus, $\alpha_2/\alpha_4$, i.e.\ the tunneling current of the steady state
relative to its total particle number, is positive describing a flux from site
2, where particles are injected, to site 1, where particles are removed.
It increases with the strength of the in- and outcoupling $\gamma$.
For large particle numbers $N_0$ this increase is approximately linear.
The quantity $\alpha_3/\alpha_4$ is the imbalance of the particles in the
steady state relative to the total particle number.
For $N_0 \gg \gamma^2/4J^2$ it is negligible, i.e., the expectation value
of the particle number at the two sites is approximately the same.

To discuss the properties of the steady state in more detail the eigenvalues
and eigenvectors of the reduced single-particle density matrix $\sigma_{jk} =
\mean{\ha_k^\dagger \ha_j}$ are calculated~\cite{Yang62a}.
The two eigenvalues are determined by the purity, $\lambda_{1/2} =
\frac{1}{2}(1\pm\sqrt{P})$, and the corresponding eigenvectors read
\begin{align}
  \bm{u}_{1/2} = \frac{1}{\sqrt{2}}
  \begin{pmatrix}
    \left(1 \mp \frac{\gamma_-}{\sqrt{4J^2+\gamma_-^2}}\right)^{1/2}
    \eul^{\pm i \pi/2}\\
    \left(1 \pm \frac{\gamma_-}{\sqrt{4J^2+\gamma_-^2}}\right)^{1/2}
  \end{pmatrix}.
\end{align}
The elements of the two eigenvectors can be interpreted as coefficients of a
single-particle state.
Of course, the similarity of the two eigenvectors stems from the fact that they
are orthogonal since the single-particle density matrix is Hermitian.
For $\gamma_- \to 0$ the two components of both eigenvectors are equal up to a
phase, and for increasing values they diverge approximately linearly since
$\gamma_- \ll 1$.
In the limit of large particle numbers $\gamma_-$ can be neglected and the
eigenvectors are given by
\begin{align}
  \bm{u}_{1/2} \approx \frac{1}{\sqrt{2}}
  \begin{pmatrix}
    \eul^{\pm i \pi/2}\\
    1
  \end{pmatrix}.
\end{align}
Since the tunneling current between the two sites is given by $j_{2\to1} = 2 J
r_1 r_2 \sin(\beta_1-\beta_2)$ (with $r_j \exp(i \beta_j)$ being the $j$th
component of $\bm{u}_{1/2}$) the approximated expression for $\bm{u}_1$
describes a single-particle state with maximum tunneling current from site 2 to
site 1 while $\bm{u}_2$ has a maximum current in the opposite direction.

Due to the incoupling of particles at site 2 and the outcoupling at site 1, a
compensating tunneling current from 2 to 1 is required for the steady state.
Because the eigenvalue to the eigenstate $\bm{u}_1$, which has a tunneling
current from 2 to 1, is larger than the other eigenvalue an effective current
from site 2 to 1 is achieved.
For increasing values of $\gamma$ the eigenvectors remain almost
unchanged but the eigenvalue $\lambda_1$ increases from $1/2$ to $1$ since the
purity $P$ of the steady state increases from $0$ to $1$ as can be seen in
Eq.~\eqref{eq:ss_purity_approx}.
Consequently $\lambda_2$ decreases towards $0$ since $\lambda_2 = 1 -
\lambda_1$.
This means a stronger compensating flux is produced by the change in the purity
and not by a change of the eigenvectors $\bm{u}_{1/2}$.

\subsection{Numerical results}
\label{sec:stat_ss_nonlinear}

To obtain the non-equilibrium steady state of the two-mode system with balanced
gain and loss numerically, one can solve the equation $\dt \hrho = 0$ using an
iterative method such as the loose generalized minimal residual method
(LGMRES)~\cite{Baker05a}.
The crucial parameter for the numerical cost is the dimension of the Fock
basis, which is determined by the maximum amount of particles at each site.
To obtain results in a reasonable time span the dimension of the Fock basis of
a state vector at a single site is chosen to be $25$, limiting the maximum
amount of particles at each site to $24$.
It is necessary to ensure that the contributions of states close to this limit
are very small since only then it is a good approximation to truncate the
basis.
For time evolutions this can be achieved by choosing the initial amount of
particles $N_0$ small enough.
However, when searching for the steady state, the particle number is not known
in advance but instead is obtained as the result of the calculation for each
set of the system's parameters such as the strength of the gain and loss or the
particle interaction.

Therefore, one has to keep in mind that in the following the parameters $N_0$
and $g$ are not the particle number and the macroscopic interaction strength of
the steady state.
Instead, the steady state is reachable by an initial state with $N_0$ particles
and interaction strength $g=U(N_0-1)$.
The initial particle number $N_0$ only determines the ratio of $\gamma_\gain$
and $\gamma_\loss$.
Since the particle number of the steady state $n$ differs, so does the value of
$g=U(n-1)$, whereas the interaction strength between two particles $U$ is held
constant.

\begin{figure}
  \centering
  \includegraphics{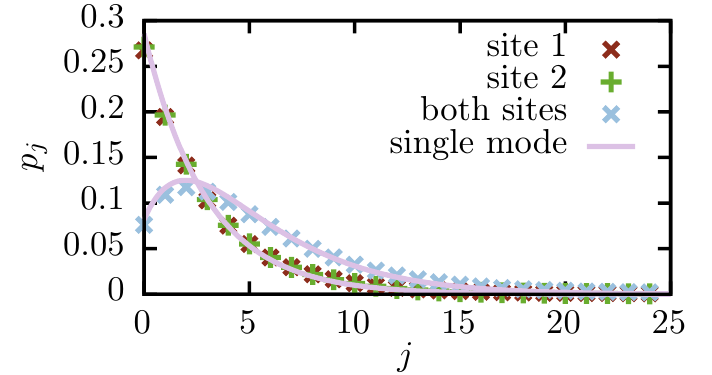}
  \caption{%
    The probabilities of finding $j$ particles at site 1, site 2 and at both
    sites obtained via the iterative approach (data points) are in very good
    agreement with the probabilities of the analytically solvable single-mode
    approach (solid lines) given by Eq.~\eqref{eq:steadySingleModeProbs} and
    Eq.~\eqref{eq:steadyStateSingleModeCombinedProbs}, respectively.
    The parameters used are $N_0=5$, $g=0.5$ and $\gamma=0.5$.
  }
  \label{fig:steadyStateManyParticle}
\end{figure}

To analyze the statistical properties of the steady state the probability of
finding a certain amount of particles at site 1 and at site 2 is shown in
Fig.~\ref{fig:steadyStateManyParticle} for $N_0=5$, $g=0.5$ and $\gamma=0.5$.
The probabilities at the two sites are almost identical and can be compared
with the behavior found in the case of gain and loss at a single site described
by
\begin{align}
  \dt \hrho_\mrm{single} =
  &-\frac{\gamma_\loss}{2} \left( \ha^\dagger \ha \hrho + \hrho
  \ha^\dagger \ha - 2 \ha \hrho \ha^\dagger \right)
  \notag\\
  &-\frac{\gamma_\gain}{2} \left( \ha \ha^\dagger \hrho + \hrho
  \ha \ha^\dagger - 2 \ha^\dagger \hrho \ha \right).
\end{align}
A study of this master equation in the context of quantum optics can be found,
e.g., in~\cite{Walls08a}.
Using the ansatz
\begin{equation}
  \hrho_\mrm{single} = \sum_{j=0}^{\infty} p_j \ket{j}\bra{j}
\end{equation}
shows that the steady state of this system is given by
\begin{equation}
  p_j = (1-\xi) \xi^j,
  \qquad
  \xi = \frac{\gamma_\gain}{\gamma_\loss}.
  \label{eq:steadySingleModeProbs}
\end{equation}

By inserting the condition for balanced gain and loss~\eqref{eq:gainlossratio}
into the probability of finding $j$ particles at the single
site~\eqref{eq:steadySingleModeProbs}, one obtains the solid line in
Fig.~\ref{fig:steadyStateManyParticle}, which lies almost perfectly on top
of the probabilities found in the two-mode system.

Furthermore, Fig.~\ref{fig:steadyStateManyParticle} shows the probability of
finding a certain amount of particles at both sites.
This quantity cannot be obtained from the probabilities at the single sites
since they might be correlated.
To check if the correlations are important for the probabilities of finding
particles at both sites, a separable state consisting of the steady state of
the single-mode calculation is constructed,
\begin{equation}
  \hrho = \hrho_\mrm{single}^{(1)} \otimes \hrho_\mrm{single}^{(2)}
  = \sum_{j,k=0}^{\infty} p_j p_k \ket{j, k}\bra{j, k},
  \label{eq:separableSteadyState}
\end{equation}
with $p_j$ given by Eq.~\eqref{eq:steadySingleModeProbs}.
The probability of finding $j$ particles in the two-mode system is the
expectation value of the operator $\hat{X}_j = \sum_{k=0}^{j} \ket{k,
j-k}\bra{k, j-k}$, which yields
\begin{equation}
  \mean{\hat{X}_j} = \sum_{k=0}^{j} p_k p_{j-k} = (1-\xi)^2 \xi^j (j+1).
  \label{eq:steadyStateSingleModeCombinedProbs}
\end{equation}
Comparing this probability distribution with the numerically obtained
probabilities shows a very good agreement.
Only for small particle numbers less than $5$, one obtains a visible
discrepancy between the two approaches, and the probabilities given by
Eq.~\eqref{eq:steadyStateSingleModeCombinedProbs} are slightly bigger.
However, this discrepancy does not arise due to correlations but instead is
mainly a consequence of the slightly larger probabilities at each site.

This shows that for the probability distribution of finding a certain amount of
particles in the system, the steady state of the two-mode system with balanced
gain and loss is very similar to a product of the steady states of the
single-mode calculation.
However, when looking at other observables there are fundamental differences.
In particular the separable state~\eqref{eq:separableSteadyState} does not have
any correlations between the two sites and as a result both $s_x$ and $s_y$
vanish which is not true for the actual steady state of the two-mode system.

For larger particle numbers the iterative approach to find the steady state is
no longer feasible since it becomes numerically too costly.
To overcome this limitation the Bogoliubov backreaction method is used.
As shown in~\cite{Dast16b} this method is a very good approximation of the
actual dynamics for a limited time span.
An approximation of the first- and second-order moments of the steady state is
obtained by a root search of the equations of motion
\begin{equation}
  \dt s_{x,y,z} = \dt n = \dt \Delta_{jk} = 0,
  \quad j,k \in \{x,y,z,n\},
  \label{eq:SSRootSearch}
\end{equation}
which is solved numerically.
Note that in general the Bogoliubov backreaction method is not able to produce
accurate results for the dynamics of an arbitrary initial state up to the time
at which it reaches the steady state since this exceeds the time span in which
this method is reliable.
However, it can be expected that the steady state obtained via the root
search is nevertheless a good approximation because the short-time behavior of
the steady state itself should be well captured.
By comparing the results of the Bogoliubov backreaction method with the results
obtained by the iterative approach, this assumption was confirmed.

\begin{figure}
  \centering
  \includegraphics[width=\columnwidth]{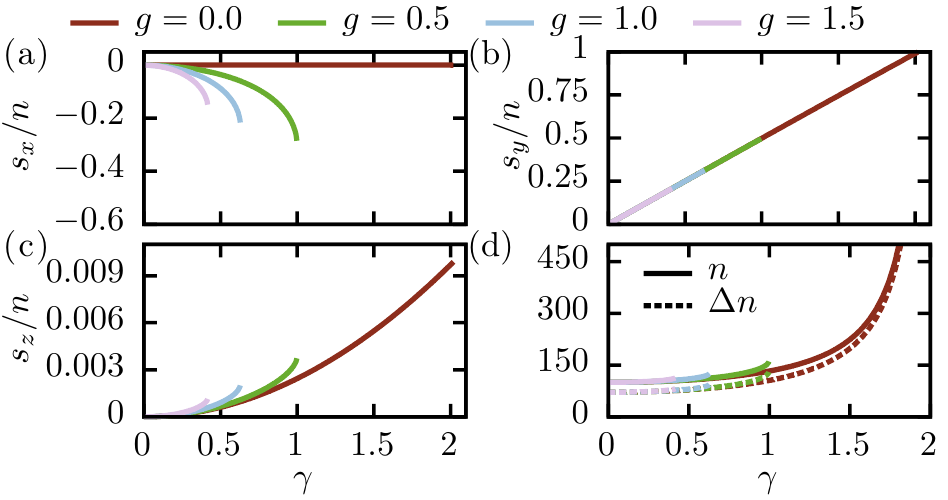}
  \caption{%
    The first-order moments $s_{x,y,z}$ and $n$ of the non-equilibrium steady
    state for the initial particle number $N_0=100$ and different values of the
    interaction strength $g$.
    The results are obtained by a root search of the equations of motion of
    the Bogoliubov backreaction method.
    In (d) the uncertainty of the particle number defined in
    Eq.~\eqref{eq:uncertainityN} is shown along with the expectation value.
  }
  \label{fig:steadyStateConstUComponents}
\end{figure}

The first-order moments in Bloch representation of the steady state for an
initial particle number $N_0=100$ are shown in
Fig.~\ref{fig:steadyStateConstUComponents} for different values of the
interaction strength $g$.
For comparison the analytically obtained solutions in the non-interacting limit
are plotted.
One immediately recognizes that the critical value of $\gamma$ at which the
steady state ceases to exist becomes considerably smaller for stronger
interactions $g$.

Another difference is that the component $s_x$, which is equal to zero for the
steady state in the non-interacting limit, becomes negative with absolute
values that are almost as large as those of $s_y$.
By contrast, the values of $s_y$ are nearly independent of the interaction
strength $g$ and are exclusively determined by the strength of the in- and
outcoupling $\gamma$.
Since $s_y$ is the particle flux from site 2 to site 1, which compensates
the outcoupling at site 1 and the incoupling at site 2, this behavior is
intuitively understandable.
The component $s_z$ of the steady state behaves similarly to the
non-interacting limit.
It is close to zero, which means that the steady state is almost balanced,
i.e., the particles are equally distributed at both sites.
With interaction the total particle number of the steady state does not reach
values as large as in the non-interacting limit.
This behavior is in agreement with the previous finding that the steady state
vanishes earlier for stronger interactions since larger particle numbers imply
a stronger macroscopic interaction $g=U(n-1)$.
To emphasize that the steady state does not have a definite particle number,
but instead there is a broad probability distribution, the uncertainty of the
expectation value of the total particle number,
\begin{equation}
  \Delta n = \sqrt{\mean{\hn^2} - \mean{\hn}^2},
  \label{eq:uncertainityN}
\end{equation}
is plotted alongside the particle number in
Fig.~\ref{fig:steadyStateConstUComponents}(d).
This shows that the uncertainty of the particle number is almost as large as
the expectation value itself.

Note that at $\gamma = 0$ the steady state is not well-defined since in this
case an infinite amount of stationary solutions exist.
Thus, the results shown in Fig.~\ref{fig:steadyStateConstUComponents} and in
the following are strictly only valid for $\gamma > 0$.

\begin{figure}
  \centering
  \includegraphics[width=\columnwidth]{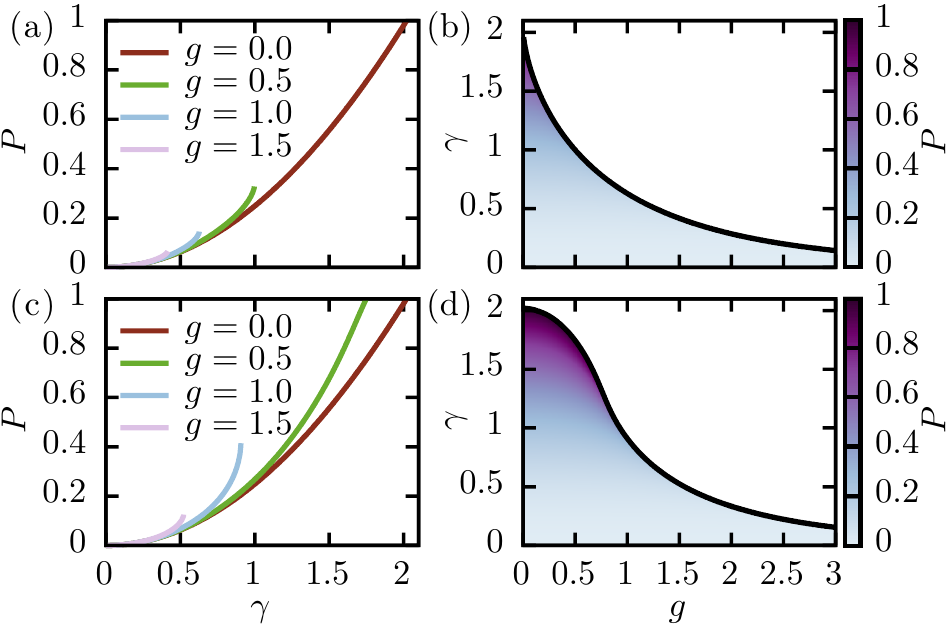}
  \caption{%
    (a) The purity of the steady state as a function of the gain-loss parameter
    $\gamma$ for different values of the interaction strength $g$.
    (b) For the parameter values in the colored area a steady state exists.
    Both the critical value of $\gamma$ where the steady state vanishes and its
    purity decrease rapidly for stronger interactions $g$.
    (c) Same as (a) but the macroscopic interaction strength $g$ is held
    constant using the replacement~\eqref{eq:SSConstG}.
    (d) With $g$ held constant the critical value of $\gamma$ at which the
    steady state vanishes decreases much slower as compared with (b) and, as a
    result, pure steady states are achieved in the presence of interaction.
    For all calculations $N_0=100$ was used.
  }
  \label{steadyStatePurity}
\end{figure}

The purity of the steady state is shown in
Fig.~\ref{steadyStatePurity}(a) for different interaction strengths.
In the non-interacting limit the steady state is perfectly pure, i.e.\ $P=1$,
at the critical value of $\gamma$ at which it vanishes.
With interaction, however, the steady state disappears at substantially smaller
purities.

Calculating the eigenvectors of the reduced single-particle density matrix
shows that similar to the non-interacting limit all absolute values of the
components stay close to $1/\sqrt{2}$.
A slightly different behavior is found for the relative phase.
In the case $g=0$ the relative phase is a constant, whereas with
interaction the phase increases with the gain-loss parameter $\gamma$.
This means that the tunneling current from site 2 to 1 of the single-particle
state to the larger eigenvalue decreases.
Equally the tunneling current of the second eigenvector in the other direction
decreases.
The complete tunneling current $s_y$, which must compensate the in- and
outcoupling of particles, however, increases with $\gamma$ since the steady
state becomes more pure and the eigenvalues of the single-particle density
matrix are given by $\lambda_{1,2} = \frac{1}{2} (1 \pm \sqrt{P})$.
Thus, as in the non-interacting case, the increasing compensating current is
generated by the steady state becoming more pure and not by a change in the
eigenvectors of the single-particle density matrix.

To obtain an overview of the parameter values in which the steady state exists,
the root search is repeated for various values of $g$ and $\gamma$ and the
result is plotted in Fig.~\ref{steadyStatePurity}(b).
The steady state exists for the parameter values in the colored area below the
solid black line and the color indicates its purity.
It can be seen that the critical value of $\gamma$ at which the steady state
vanishes decreases rapidly for stronger particle interactions $g$ and so does
the purity.

Up to now the steady state that is reachable in the two-mode system with
balanced gain and loss by an initial state with $N_0$ particles and macroscopic
interaction strength $g = U(N_0-1)$ was calculated.
However, the macroscopic interaction of the steady state is then given by
$g_\mathrm{steady} = g (n_\mathrm{steady}-1)/(N_0-1)$ with the particle number
of the steady state $n_\mathrm{steady}$.
Since the particle number $n_\mathrm{steady}$ can become very large close to
the critical value of $\gamma$ where the steady state vanishes, this has a
marked effect on the system.

A different approach is to ask whether a steady state exists for a specific
macroscopic interaction $g$.
To achieve this, the parameter $U$ is replaced by
\begin{equation}
  U(t) = \frac{g}{n(t)-1}
  \label{eq:SSConstG}
\end{equation}
in the equations of motion of the Bogoliubov backreaction method, with $n(t)$
being the time-dependent particle number.
The steady state is then again found via the root search defined in
Eq.~\eqref{eq:SSRootSearch}.
Note that for this root search the ratio of the gain and loss contributions is
still given by Eq.~\eqref{eq:gainlossratio}, i.e., they are held constant.

The purity obtained in this manner is shown in
Fig.~\ref{steadyStatePurity}(c) for different values of $g$.
The crucial difference to the results obtained previously is that the steady
state now exists for larger values of $\gamma$ resulting in a purer steady
state.
In particular, a perfectly pure steady state is also possible in presence of
interaction as can be seen for $g = 0.5$.

Calculating the steady state for various values of $\gamma$ and $g$ yields the
results shown in Fig.~\ref{steadyStatePurity}(d).
The main difference to Fig.~\ref{steadyStatePurity}(b) is that the
critical value of $\gamma$ at which the steady state vanishes decreases much
slower for small values of $g$.
Consequently, purities equal or close to one are achieved in the presence of
interaction.
In the parameter regime $\gamma \in [1.5,2]$ and $g \in [0,0.75]$ the steady
state is almost pure, thus, there exists a pure condensate which is actually
stationary.
For stronger interaction strengths $g$ the results obtained by the two
approaches become similar.

\section{Conclusion}
\label{sec:conclusion}

In this work the role of stationary states in the many-particle description of
a Bose-Einstein condensate with balanced gain and loss was studied.

It was shown that in the many-particle description the outstanding property of
the $\PT$-symmetric states of the mean-field description is not that they are
stationary but instead that they show only weak oscillations.
Without interaction two distinguished trajectories exist in the vicinity of
both $\PT$-symmetric states, which show no oscillations at all.
For large particle numbers these non-oscillatory states become equal to the
$\PT$-symmetric mean-field states.
By looking at their time evolution using the Bloch sphere formalism, one
immediately recognizes their special role since all other trajectories encircle
one of the two non-oscillating trajectories.
The fact that in the mean-field limit the $\PT$-symmetric stationary states
appear as elliptic fixed points which are encircled by oscillating
trajectories motivates the interpretation of the non-oscillatory states as the
many-particle equivalent of the $\PT$-symmetric states.

Furthermore, the non-equilibrium steady state of the master equation with
balanced gain and loss was calculated.
In the non-interacting limit the steady state was obtained analytically.
The calculation showed that without coupling to the environment the purity of
the steady state is minimum but it increases for stronger values of the
gain-loss parameter, and eventually the state becomes completely pure at a
critical value.
In this parameter regime the steady state is an attractor that every trajectory
finally reaches in the limit $t \to \infty$.

With interaction the steady state was calculated numerically showing a similar
behavior.
However, the steady state now only exists for much smaller gain-loss
contributions and it only reaches substantially smaller purities.
The reason for this behavior is that the particle number of the steady state
increases with the gain-loss parameter, and thus the influence of the
interaction becomes very strong.
By fixing the macroscopic interaction strength of the steady state, it was
shown that a pure steady state can also be reached in the presence of
interaction.
In fact there is a whole parameter regime in which a pure stationary condensate
exists.
Finally, calculating the eigenvectors of the single-particle density matrix of
the non-equilibrium steady state showed that the increase of the effective
current is not generated by a change in the eigenvectors but instead by the
steady state becoming more pure, thus, one eigenvector becoming more important
than the other one.

Compared with previous studies~\cite{Witthaut09a, Witthaut11a} where a
Bose-Einstein condensate subject to localized particle loss was investigated,
we were able to find a true steady state in the presence of particle
interaction, whereas in those systems only an exponentially decaying
quasi-steady state exists in the non-interacting limit and in an approximation
for weak interactions.
Furthermore the purification process driven by particle gain and
loss~\cite{Dast16a, Dast16b} as applied in this study is fundamentally
different from that of a system with only loss~\cite{Witthaut09a, Witthaut11a}.


\begin{thebibliography}{43}%
\makeatletter
\providecommand \@ifxundefined [1]{%
 \@ifx{#1\undefined}
}%
\providecommand \@ifnum [1]{%
 \ifnum #1\expandafter \@firstoftwo
 \else \expandafter \@secondoftwo
 \fi
}%
\providecommand \@ifx [1]{%
 \ifx #1\expandafter \@firstoftwo
 \else \expandafter \@secondoftwo
 \fi
}%
\providecommand \natexlab [1]{#1}%
\providecommand \enquote  [1]{``#1''}%
\providecommand \bibnamefont  [1]{#1}%
\providecommand \bibfnamefont [1]{#1}%
\providecommand \citenamefont [1]{#1}%
\providecommand \href@noop [0]{\@secondoftwo}%
\providecommand \href [0]{\begingroup \@sanitize@url \@href}%
\providecommand \@href[1]{\@@startlink{#1}\@@href}%
\providecommand \@@href[1]{\endgroup#1\@@endlink}%
\providecommand \@sanitize@url [0]{\catcode `\\12\catcode `\$12\catcode
  `\&12\catcode `\#12\catcode `\^12\catcode `\_12\catcode `\%12\relax}%
\providecommand \@@startlink[1]{}%
\providecommand \@@endlink[0]{}%
\providecommand \url  [0]{\begingroup\@sanitize@url \@url }%
\providecommand \@url [1]{\endgroup\@href {#1}{\urlprefix }}%
\providecommand \urlprefix  [0]{URL }%
\providecommand \Eprint [0]{\href }%
\providecommand \doibase [0]{http://dx.doi.org/}%
\providecommand \selectlanguage [0]{\@gobble}%
\providecommand \bibinfo  [0]{\@secondoftwo}%
\providecommand \bibfield  [0]{\@secondoftwo}%
\providecommand \translation [1]{[#1]}%
\providecommand \BibitemOpen [0]{}%
\providecommand \bibitemStop [0]{}%
\providecommand \bibitemNoStop [0]{.\EOS\space}%
\providecommand \EOS [0]{\spacefactor3000\relax}%
\providecommand \BibitemShut  [1]{\csname bibitem#1\endcsname}%
\let\auto@bib@innerbib\@empty
\bibitem [{\citenamefont {Moiseyev}(2011)}]{Moiseyev11a}%
  \BibitemOpen
  \bibfield  {author} {\bibinfo {author} {\bibfnamefont {N.}~\bibnamefont
  {Moiseyev}},\ }\href@noop {} {\emph {\bibinfo {title} {{Non-Hermitian Quantum
  Mechanics}}}}\ (\bibinfo  {publisher} {Cambridge University Press},\ \bibinfo
  {address} {Cambridge},\ \bibinfo {year} {2011})\BibitemShut {NoStop}%
\bibitem [{\citenamefont {Bender}\ and\ \citenamefont
  {Boettcher}(1998)}]{Bender98a}%
  \BibitemOpen
  \bibfield  {author} {\bibinfo {author} {\bibfnamefont {C.~M.}\ \bibnamefont
  {Bender}}\ and\ \bibinfo {author} {\bibfnamefont {S.}~\bibnamefont
  {Boettcher}},\ }\href {\doibase 10.1103/PhysRevLett.80.5243} {\bibfield
  {journal} {\bibinfo  {journal} {Phys. Rev. Lett.}\ }\textbf {\bibinfo
  {volume} {80}},\ \bibinfo {pages} {5243} (\bibinfo {year}
  {1998})}\BibitemShut {NoStop}%
\bibitem [{\citenamefont {Bender}\ \emph {et~al.}(2002)\citenamefont {Bender},
  \citenamefont {Brody},\ and\ \citenamefont {Jones}}]{Bender02a}%
  \BibitemOpen
  \bibfield  {author} {\bibinfo {author} {\bibfnamefont {C.~M.}\ \bibnamefont
  {Bender}}, \bibinfo {author} {\bibfnamefont {D.~C.}\ \bibnamefont {Brody}}, \
  and\ \bibinfo {author} {\bibfnamefont {H.~F.}\ \bibnamefont {Jones}},\ }\href
  {\doibase 10.1103/PhysRevLett.89.270401} {\bibfield  {journal} {\bibinfo
  {journal} {Phys. Rev. Lett.}\ }\textbf {\bibinfo {volume} {89}},\ \bibinfo
  {pages} {270401} (\bibinfo {year} {2002})}\BibitemShut {NoStop}%
\bibitem [{\citenamefont {Kagan}\ \emph {et~al.}(1998)\citenamefont {Kagan},
  \citenamefont {Muryshev},\ and\ \citenamefont {Shlyapnikov}}]{Kagan98a}%
  \BibitemOpen
  \bibfield  {author} {\bibinfo {author} {\bibfnamefont {Y.}~\bibnamefont
  {Kagan}}, \bibinfo {author} {\bibfnamefont {A.~E.}\ \bibnamefont {Muryshev}},
  \ and\ \bibinfo {author} {\bibfnamefont {G.~V.}\ \bibnamefont
  {Shlyapnikov}},\ }\href {\doibase 10.1103/PhysRevLett.81.933} {\bibfield
  {journal} {\bibinfo  {journal} {Phys. Rev. Lett.}\ }\textbf {\bibinfo
  {volume} {81}},\ \bibinfo {pages} {933} (\bibinfo {year} {1998})}\BibitemShut
  {NoStop}%
\bibitem [{\citenamefont {Gericke}\ \emph {et~al.}(2008)\citenamefont
  {Gericke}, \citenamefont {Wurtz}, \citenamefont {Reitz}, \citenamefont
  {Langen},\ and\ \citenamefont {Ott}}]{Gericke08a}%
  \BibitemOpen
  \bibfield  {author} {\bibinfo {author} {\bibfnamefont {T.}~\bibnamefont
  {Gericke}}, \bibinfo {author} {\bibfnamefont {P.}~\bibnamefont {Wurtz}},
  \bibinfo {author} {\bibfnamefont {D.}~\bibnamefont {Reitz}}, \bibinfo
  {author} {\bibfnamefont {T.}~\bibnamefont {Langen}}, \ and\ \bibinfo {author}
  {\bibfnamefont {H.}~\bibnamefont {Ott}},\ }\href {\doibase 10.1038/nphys1102}
  {\bibfield  {journal} {\bibinfo  {journal} {Nat. Phys.}\ }\textbf {\bibinfo
  {volume} {4}},\ \bibinfo {pages} {949} (\bibinfo {year} {2008})}\BibitemShut
  {NoStop}%
\bibitem [{\citenamefont {W\"urtz}\ \emph {et~al.}(2009)\citenamefont
  {W\"urtz}, \citenamefont {Langen}, \citenamefont {Gericke}, \citenamefont
  {Koglbauer},\ and\ \citenamefont {Ott}}]{Wurtz09a}%
  \BibitemOpen
  \bibfield  {author} {\bibinfo {author} {\bibfnamefont {P.}~\bibnamefont
  {W\"urtz}}, \bibinfo {author} {\bibfnamefont {T.}~\bibnamefont {Langen}},
  \bibinfo {author} {\bibfnamefont {T.}~\bibnamefont {Gericke}}, \bibinfo
  {author} {\bibfnamefont {A.}~\bibnamefont {Koglbauer}}, \ and\ \bibinfo
  {author} {\bibfnamefont {H.}~\bibnamefont {Ott}},\ }\href {\doibase
  10.1103/PhysRevLett.103.080404} {\bibfield  {journal} {\bibinfo  {journal}
  {Phys. Rev. Lett.}\ }\textbf {\bibinfo {volume} {103}},\ \bibinfo {pages}
  {080404} (\bibinfo {year} {2009})}\BibitemShut {NoStop}%
\bibitem [{\citenamefont {Barontini}\ \emph {et~al.}(2013)\citenamefont
  {Barontini}, \citenamefont {Labouvie}, \citenamefont {Stubenrauch},
  \citenamefont {Vogler}, \citenamefont {Guarrera},\ and\ \citenamefont
  {Ott}}]{Barontini13a}%
  \BibitemOpen
  \bibfield  {author} {\bibinfo {author} {\bibfnamefont {G.}~\bibnamefont
  {Barontini}}, \bibinfo {author} {\bibfnamefont {R.}~\bibnamefont {Labouvie}},
  \bibinfo {author} {\bibfnamefont {F.}~\bibnamefont {Stubenrauch}}, \bibinfo
  {author} {\bibfnamefont {A.}~\bibnamefont {Vogler}}, \bibinfo {author}
  {\bibfnamefont {V.}~\bibnamefont {Guarrera}}, \ and\ \bibinfo {author}
  {\bibfnamefont {H.}~\bibnamefont {Ott}},\ }\href {\doibase
  10.1103/PhysRevLett.110.035302} {\bibfield  {journal} {\bibinfo  {journal}
  {Phys. Rev. Lett.}\ }\textbf {\bibinfo {volume} {110}},\ \bibinfo {pages}
  {035302} (\bibinfo {year} {2013})}\BibitemShut {NoStop}%
\bibitem [{\citenamefont {Robins}\ \emph {et~al.}(2008)\citenamefont {Robins},
  \citenamefont {Figl}, \citenamefont {Jeppesen}, \citenamefont {Dennis},\ and\
  \citenamefont {Close}}]{Robins08a}%
  \BibitemOpen
  \bibfield  {author} {\bibinfo {author} {\bibfnamefont {N.~P.}\ \bibnamefont
  {Robins}}, \bibinfo {author} {\bibfnamefont {C.}~\bibnamefont {Figl}},
  \bibinfo {author} {\bibfnamefont {M.}~\bibnamefont {Jeppesen}}, \bibinfo
  {author} {\bibfnamefont {G.~R.}\ \bibnamefont {Dennis}}, \ and\ \bibinfo
  {author} {\bibfnamefont {J.~D.}\ \bibnamefont {Close}},\ }\href {\doibase
  10.1038/nphys1027} {\bibfield  {journal} {\bibinfo  {journal} {Nat. Phys.}\
  }\textbf {\bibinfo {volume} {4}},\ \bibinfo {pages} {731} (\bibinfo {year}
  {2008})}\BibitemShut {NoStop}%
\bibitem [{\citenamefont {D\"oring}\ \emph {et~al.}(2009)\citenamefont
  {D\"oring}, \citenamefont {Dennis}, \citenamefont {Robins}, \citenamefont
  {Jeppesen}, \citenamefont {Figl}, \citenamefont {Hope},\ and\ \citenamefont
  {Close}}]{Doring09a}%
  \BibitemOpen
  \bibfield  {author} {\bibinfo {author} {\bibfnamefont {D.}~\bibnamefont
  {D\"oring}}, \bibinfo {author} {\bibfnamefont {G.~R.}\ \bibnamefont
  {Dennis}}, \bibinfo {author} {\bibfnamefont {N.~P.}\ \bibnamefont {Robins}},
  \bibinfo {author} {\bibfnamefont {M.}~\bibnamefont {Jeppesen}}, \bibinfo
  {author} {\bibfnamefont {C.}~\bibnamefont {Figl}}, \bibinfo {author}
  {\bibfnamefont {J.~J.}\ \bibnamefont {Hope}}, \ and\ \bibinfo {author}
  {\bibfnamefont {J.~D.}\ \bibnamefont {Close}},\ }\href {\doibase
  10.1103/PhysRevA.79.063630} {\bibfield  {journal} {\bibinfo  {journal} {Phys.
  Rev. A}\ }\textbf {\bibinfo {volume} {79}},\ \bibinfo {pages} {063630}
  (\bibinfo {year} {2009})}\BibitemShut {NoStop}%
\bibitem [{\citenamefont {Schneble}\ \emph {et~al.}(2004)\citenamefont
  {Schneble}, \citenamefont {Campbell}, \citenamefont {Streed}, \citenamefont
  {Boyd}, \citenamefont {Pritchard},\ and\ \citenamefont
  {Ketterle}}]{Schneble04a}%
  \BibitemOpen
  \bibfield  {author} {\bibinfo {author} {\bibfnamefont {D.}~\bibnamefont
  {Schneble}}, \bibinfo {author} {\bibfnamefont {G.~K.}\ \bibnamefont
  {Campbell}}, \bibinfo {author} {\bibfnamefont {E.~W.}\ \bibnamefont
  {Streed}}, \bibinfo {author} {\bibfnamefont {M.}~\bibnamefont {Boyd}},
  \bibinfo {author} {\bibfnamefont {D.~E.}\ \bibnamefont {Pritchard}}, \ and\
  \bibinfo {author} {\bibfnamefont {W.}~\bibnamefont {Ketterle}},\ }\href
  {\doibase 10.1103/PhysRevA.69.041601} {\bibfield  {journal} {\bibinfo
  {journal} {Phys. Rev. A}\ }\textbf {\bibinfo {volume} {69}},\ \bibinfo
  {pages} {041601} (\bibinfo {year} {2004})}\BibitemShut {NoStop}%
\bibitem [{\citenamefont {Yoshikawa}\ \emph {et~al.}(2004)\citenamefont
  {Yoshikawa}, \citenamefont {Sugiura}, \citenamefont {Torii},\ and\
  \citenamefont {Kuga}}]{Yoshikawa04a}%
  \BibitemOpen
  \bibfield  {author} {\bibinfo {author} {\bibfnamefont {Y.}~\bibnamefont
  {Yoshikawa}}, \bibinfo {author} {\bibfnamefont {T.}~\bibnamefont {Sugiura}},
  \bibinfo {author} {\bibfnamefont {Y.}~\bibnamefont {Torii}}, \ and\ \bibinfo
  {author} {\bibfnamefont {T.}~\bibnamefont {Kuga}},\ }\href {\doibase
  10.1103/PhysRevA.69.041603} {\bibfield  {journal} {\bibinfo  {journal} {Phys.
  Rev. A}\ }\textbf {\bibinfo {volume} {69}},\ \bibinfo {pages} {041603}
  (\bibinfo {year} {2004})}\BibitemShut {NoStop}%
\bibitem [{\citenamefont {Graefe}(2012)}]{Graefe12a}%
  \BibitemOpen
  \bibfield  {author} {\bibinfo {author} {\bibfnamefont {E.-M.}\ \bibnamefont
  {Graefe}},\ }\href {\doibase 10.1088/1751-8113/45/44/444015} {\bibfield
  {journal} {\bibinfo  {journal} {J. Phys. A}\ }\textbf {\bibinfo {volume}
  {45}},\ \bibinfo {pages} {444015} (\bibinfo {year} {2012})}\BibitemShut
  {NoStop}%
\bibitem [{\citenamefont {Haag}\ \emph {et~al.}(2015)\citenamefont {Haag},
  \citenamefont {Dast}, \citenamefont {Cartarius},\ and\ \citenamefont
  {Wunner}}]{Haag15a}%
  \BibitemOpen
  \bibfield  {author} {\bibinfo {author} {\bibfnamefont {D.}~\bibnamefont
  {Haag}}, \bibinfo {author} {\bibfnamefont {D.}~\bibnamefont {Dast}}, \bibinfo
  {author} {\bibfnamefont {H.}~\bibnamefont {Cartarius}}, \ and\ \bibinfo
  {author} {\bibfnamefont {G.}~\bibnamefont {Wunner}},\ }\href {\doibase
  10.1103/PhysRevA.92.053627} {\bibfield  {journal} {\bibinfo  {journal} {Phys.
  Rev. A}\ }\textbf {\bibinfo {volume} {92}},\ \bibinfo {pages} {053627}
  (\bibinfo {year} {2015})}\BibitemShut {NoStop}%
\bibitem [{\citenamefont {Cartarius}\ and\ \citenamefont
  {Wunner}(2012)}]{Cartarius12b}%
  \BibitemOpen
  \bibfield  {author} {\bibinfo {author} {\bibfnamefont {H.}~\bibnamefont
  {Cartarius}}\ and\ \bibinfo {author} {\bibfnamefont {G.}~\bibnamefont
  {Wunner}},\ }\href {\doibase 10.1103/physreva.86.013612} {\bibfield
  {journal} {\bibinfo  {journal} {Phys. Rev. A}\ }\textbf {\bibinfo {volume}
  {86}},\ \bibinfo {pages} {013612} (\bibinfo {year} {2012})}\BibitemShut
  {NoStop}%
\bibitem [{\citenamefont {Cartarius}\ \emph {et~al.}(2012)\citenamefont
  {Cartarius}, \citenamefont {Haag}, \citenamefont {Dast},\ and\ \citenamefont
  {Wunner}}]{Cartarius12a}%
  \BibitemOpen
  \bibfield  {author} {\bibinfo {author} {\bibfnamefont {H.}~\bibnamefont
  {Cartarius}}, \bibinfo {author} {\bibfnamefont {D.}~\bibnamefont {Haag}},
  \bibinfo {author} {\bibfnamefont {D.}~\bibnamefont {Dast}}, \ and\ \bibinfo
  {author} {\bibfnamefont {G.}~\bibnamefont {Wunner}},\ }\href {\doibase
  10.1088/1751-8113/45/44/444008} {\bibfield  {journal} {\bibinfo  {journal}
  {J. Phys. A}\ }\textbf {\bibinfo {volume} {45}},\ \bibinfo {pages} {444008}
  (\bibinfo {year} {2012})}\BibitemShut {NoStop}%
\bibitem [{\citenamefont {Dast}\ \emph
  {et~al.}(2013{\natexlab{a}})\citenamefont {Dast}, \citenamefont {Haag},
  \citenamefont {Cartarius}, \citenamefont {Wunner}, \citenamefont {Eichler},\
  and\ \citenamefont {Main}}]{Dast13a}%
  \BibitemOpen
  \bibfield  {author} {\bibinfo {author} {\bibfnamefont {D.}~\bibnamefont
  {Dast}}, \bibinfo {author} {\bibfnamefont {D.}~\bibnamefont {Haag}}, \bibinfo
  {author} {\bibfnamefont {H.}~\bibnamefont {Cartarius}}, \bibinfo {author}
  {\bibfnamefont {G.}~\bibnamefont {Wunner}}, \bibinfo {author} {\bibfnamefont
  {R.}~\bibnamefont {Eichler}}, \ and\ \bibinfo {author} {\bibfnamefont
  {J.}~\bibnamefont {Main}},\ }\href {\doibase 10.1002/prop.201200080}
  {\bibfield  {journal} {\bibinfo  {journal} {Fortschr. Phys.}\ }\textbf
  {\bibinfo {volume} {61}},\ \bibinfo {pages} {124} (\bibinfo {year}
  {2013}{\natexlab{a}})}\BibitemShut {NoStop}%
\bibitem [{\citenamefont {Dast}\ \emph
  {et~al.}(2013{\natexlab{b}})\citenamefont {Dast}, \citenamefont {Haag},
  \citenamefont {Cartarius}, \citenamefont {Main},\ and\ \citenamefont
  {Wunner}}]{Dast13b}%
  \BibitemOpen
  \bibfield  {author} {\bibinfo {author} {\bibfnamefont {D.}~\bibnamefont
  {Dast}}, \bibinfo {author} {\bibfnamefont {D.}~\bibnamefont {Haag}}, \bibinfo
  {author} {\bibfnamefont {H.}~\bibnamefont {Cartarius}}, \bibinfo {author}
  {\bibfnamefont {J.}~\bibnamefont {Main}}, \ and\ \bibinfo {author}
  {\bibfnamefont {G.}~\bibnamefont {Wunner}},\ }\href {\doibase
  10.1088/1751-8113/46/37/375301} {\bibfield  {journal} {\bibinfo  {journal}
  {J. Phys. A}\ }\textbf {\bibinfo {volume} {46}},\ \bibinfo {pages} {375301}
  (\bibinfo {year} {2013}{\natexlab{b}})}\BibitemShut {NoStop}%
\bibitem [{\citenamefont {Haag}\ \emph {et~al.}(2014)\citenamefont {Haag},
  \citenamefont {Dast}, \citenamefont {L\"ohle}, \citenamefont {Cartarius},
  \citenamefont {Main},\ and\ \citenamefont {Wunner}}]{Haag14a}%
  \BibitemOpen
  \bibfield  {author} {\bibinfo {author} {\bibfnamefont {D.}~\bibnamefont
  {Haag}}, \bibinfo {author} {\bibfnamefont {D.}~\bibnamefont {Dast}}, \bibinfo
  {author} {\bibfnamefont {A.}~\bibnamefont {L\"ohle}}, \bibinfo {author}
  {\bibfnamefont {H.}~\bibnamefont {Cartarius}}, \bibinfo {author}
  {\bibfnamefont {J.}~\bibnamefont {Main}}, \ and\ \bibinfo {author}
  {\bibfnamefont {G.}~\bibnamefont {Wunner}},\ }\href {\doibase
  10.1103/PhysRevA.89.023601} {\bibfield  {journal} {\bibinfo  {journal} {Phys.
  Rev. A}\ }\textbf {\bibinfo {volume} {89}},\ \bibinfo {pages} {023601}
  (\bibinfo {year} {2014})}\BibitemShut {NoStop}%
\bibitem [{\citenamefont {Dizdarevic}\ \emph {et~al.}(2015)\citenamefont
  {Dizdarevic}, \citenamefont {Dast}, \citenamefont {Haag}, \citenamefont
  {Main}, \citenamefont {Cartarius},\ and\ \citenamefont
  {Wunner}}]{Dizdarevic15a}%
  \BibitemOpen
  \bibfield  {author} {\bibinfo {author} {\bibfnamefont {D.}~\bibnamefont
  {Dizdarevic}}, \bibinfo {author} {\bibfnamefont {D.}~\bibnamefont {Dast}},
  \bibinfo {author} {\bibfnamefont {D.}~\bibnamefont {Haag}}, \bibinfo {author}
  {\bibfnamefont {J.}~\bibnamefont {Main}}, \bibinfo {author} {\bibfnamefont
  {H.}~\bibnamefont {Cartarius}}, \ and\ \bibinfo {author} {\bibfnamefont
  {G.}~\bibnamefont {Wunner}},\ }\href {\doibase 10.1103/PhysRevA.91.033636}
  {\bibfield  {journal} {\bibinfo  {journal} {Phys. Rev. A}\ }\textbf {\bibinfo
  {volume} {91}},\ \bibinfo {pages} {033636} (\bibinfo {year}
  {2015})}\BibitemShut {NoStop}%
\bibitem [{\citenamefont {Kreibich}\ \emph {et~al.}(2013)\citenamefont
  {Kreibich}, \citenamefont {Main}, \citenamefont {Cartarius},\ and\
  \citenamefont {Wunner}}]{Kreibich13a}%
  \BibitemOpen
  \bibfield  {author} {\bibinfo {author} {\bibfnamefont {M.}~\bibnamefont
  {Kreibich}}, \bibinfo {author} {\bibfnamefont {J.}~\bibnamefont {Main}},
  \bibinfo {author} {\bibfnamefont {H.}~\bibnamefont {Cartarius}}, \ and\
  \bibinfo {author} {\bibfnamefont {G.}~\bibnamefont {Wunner}},\ }\href
  {\doibase 10.1103/PhysRevA.87.051601} {\bibfield  {journal} {\bibinfo
  {journal} {Phys. Rev. A}\ }\textbf {\bibinfo {volume} {87}},\ \bibinfo
  {pages} {051601(R)} (\bibinfo {year} {2013})}\BibitemShut {NoStop}%
\bibitem [{\citenamefont {Kreibich}\ \emph {et~al.}(2014)\citenamefont
  {Kreibich}, \citenamefont {Main}, \citenamefont {Cartarius},\ and\
  \citenamefont {Wunner}}]{Kreibich14a}%
  \BibitemOpen
  \bibfield  {author} {\bibinfo {author} {\bibfnamefont {M.}~\bibnamefont
  {Kreibich}}, \bibinfo {author} {\bibfnamefont {J.}~\bibnamefont {Main}},
  \bibinfo {author} {\bibfnamefont {H.}~\bibnamefont {Cartarius}}, \ and\
  \bibinfo {author} {\bibfnamefont {G.}~\bibnamefont {Wunner}},\ }\href
  {\doibase 10.1103/PhysRevA.90.033630} {\bibfield  {journal} {\bibinfo
  {journal} {Phys. Rev. A}\ }\textbf {\bibinfo {volume} {90}},\ \bibinfo
  {pages} {033630} (\bibinfo {year} {2014})}\BibitemShut {NoStop}%
\bibitem [{\citenamefont {Ruostekoski}\ and\ \citenamefont
  {Walls}(1998)}]{Ruostekoski98a}%
  \BibitemOpen
  \bibfield  {author} {\bibinfo {author} {\bibfnamefont {J.}~\bibnamefont
  {Ruostekoski}}\ and\ \bibinfo {author} {\bibfnamefont {D.~F.}\ \bibnamefont
  {Walls}},\ }\href {\doibase 10.1103/PhysRevA.58.R50} {\bibfield  {journal}
  {\bibinfo  {journal} {Phys. Rev. A}\ }\textbf {\bibinfo {volume} {58}},\
  \bibinfo {pages} {R50} (\bibinfo {year} {1998})}\BibitemShut {NoStop}%
\bibitem [{\citenamefont {Syassen}\ \emph {et~al.}(2008)\citenamefont
  {Syassen}, \citenamefont {Bauer}, \citenamefont {Lettner}, \citenamefont
  {Volz}, \citenamefont {Dietze}, \citenamefont {Garc{\'\i}a-Ripoll},
  \citenamefont {Cirac}, \citenamefont {Rempe},\ and\ \citenamefont
  {D{\"u}rr}}]{Syassen08a}%
  \BibitemOpen
  \bibfield  {author} {\bibinfo {author} {\bibfnamefont {N.}~\bibnamefont
  {Syassen}}, \bibinfo {author} {\bibfnamefont {D.~M.}\ \bibnamefont {Bauer}},
  \bibinfo {author} {\bibfnamefont {M.}~\bibnamefont {Lettner}}, \bibinfo
  {author} {\bibfnamefont {T.}~\bibnamefont {Volz}}, \bibinfo {author}
  {\bibfnamefont {D.}~\bibnamefont {Dietze}}, \bibinfo {author} {\bibfnamefont
  {J.~J.}\ \bibnamefont {Garc{\'\i}a-Ripoll}}, \bibinfo {author} {\bibfnamefont
  {J.~I.}\ \bibnamefont {Cirac}}, \bibinfo {author} {\bibfnamefont
  {G.}~\bibnamefont {Rempe}}, \ and\ \bibinfo {author} {\bibfnamefont
  {S.}~\bibnamefont {D{\"u}rr}},\ }\href {\doibase 10.1126/science.1155309}
  {\bibfield  {journal} {\bibinfo  {journal} {Science}\ }\textbf {\bibinfo
  {volume} {320}},\ \bibinfo {pages} {1329} (\bibinfo {year}
  {2008})}\BibitemShut {NoStop}%
\bibitem [{\citenamefont {Witthaut}\ \emph {et~al.}(2008)\citenamefont
  {Witthaut}, \citenamefont {Trimborn},\ and\ \citenamefont
  {Wimberger}}]{Witthaut08a}%
  \BibitemOpen
  \bibfield  {author} {\bibinfo {author} {\bibfnamefont {D.}~\bibnamefont
  {Witthaut}}, \bibinfo {author} {\bibfnamefont {F.}~\bibnamefont {Trimborn}},
  \ and\ \bibinfo {author} {\bibfnamefont {S.}~\bibnamefont {Wimberger}},\
  }\href {\doibase 10.1103/PhysRevLett.101.200402} {\bibfield  {journal}
  {\bibinfo  {journal} {Phys. Rev. Lett.}\ }\textbf {\bibinfo {volume} {101}},\
  \bibinfo {pages} {200402} (\bibinfo {year} {2008})}\BibitemShut {NoStop}%
\bibitem [{\citenamefont {Witthaut}\ \emph {et~al.}(2009)\citenamefont
  {Witthaut}, \citenamefont {Trimborn},\ and\ \citenamefont
  {Wimberger}}]{Witthaut09a}%
  \BibitemOpen
  \bibfield  {author} {\bibinfo {author} {\bibfnamefont {D.}~\bibnamefont
  {Witthaut}}, \bibinfo {author} {\bibfnamefont {F.}~\bibnamefont {Trimborn}},
  \ and\ \bibinfo {author} {\bibfnamefont {S.}~\bibnamefont {Wimberger}},\
  }\href {\doibase 10.1103/PhysRevA.79.033621} {\bibfield  {journal} {\bibinfo
  {journal} {Phys. Rev. A}\ }\textbf {\bibinfo {volume} {79}},\ \bibinfo
  {pages} {033621} (\bibinfo {year} {2009})}\BibitemShut {NoStop}%
\bibitem [{\citenamefont {Witthaut}\ \emph {et~al.}(2011)\citenamefont
  {Witthaut}, \citenamefont {Trimborn}, \citenamefont {Hennig}, \citenamefont
  {Kordas}, \citenamefont {Geisel},\ and\ \citenamefont
  {Wimberger}}]{Witthaut11a}%
  \BibitemOpen
  \bibfield  {author} {\bibinfo {author} {\bibfnamefont {D.}~\bibnamefont
  {Witthaut}}, \bibinfo {author} {\bibfnamefont {F.}~\bibnamefont {Trimborn}},
  \bibinfo {author} {\bibfnamefont {H.}~\bibnamefont {Hennig}}, \bibinfo
  {author} {\bibfnamefont {G.}~\bibnamefont {Kordas}}, \bibinfo {author}
  {\bibfnamefont {T.}~\bibnamefont {Geisel}}, \ and\ \bibinfo {author}
  {\bibfnamefont {S.}~\bibnamefont {Wimberger}},\ }\href {\doibase
  10.1103/PhysRevA.83.063608} {\bibfield  {journal} {\bibinfo  {journal} {Phys.
  Rev. A}\ }\textbf {\bibinfo {volume} {83}},\ \bibinfo {pages} {063608}
  (\bibinfo {year} {2011})}\BibitemShut {NoStop}%
\bibitem [{\citenamefont {Barmettler}\ and\ \citenamefont
  {Kollath}(2011)}]{Barmettler11a}%
  \BibitemOpen
  \bibfield  {author} {\bibinfo {author} {\bibfnamefont {P.}~\bibnamefont
  {Barmettler}}\ and\ \bibinfo {author} {\bibfnamefont {C.}~\bibnamefont
  {Kollath}},\ }\href {\doibase 10.1103/PhysRevA.84.041606} {\bibfield
  {journal} {\bibinfo  {journal} {Phys. Rev. A}\ }\textbf {\bibinfo {volume}
  {84}},\ \bibinfo {pages} {041606} (\bibinfo {year} {2011})}\BibitemShut
  {NoStop}%
\bibitem [{\citenamefont {Labouvie}\ \emph {et~al.}(2016)\citenamefont
  {Labouvie}, \citenamefont {Santra}, \citenamefont {Heun},\ and\ \citenamefont
  {Ott}}]{Labouvie16a}%
  \BibitemOpen
  \bibfield  {author} {\bibinfo {author} {\bibfnamefont {R.}~\bibnamefont
  {Labouvie}}, \bibinfo {author} {\bibfnamefont {B.}~\bibnamefont {Santra}},
  \bibinfo {author} {\bibfnamefont {S.}~\bibnamefont {Heun}}, \ and\ \bibinfo
  {author} {\bibfnamefont {H.}~\bibnamefont {Ott}},\ }\href {\doibase
  10.1103/PhysRevLett.116.235302} {\bibfield  {journal} {\bibinfo  {journal}
  {Phys. Rev. Lett.}\ }\textbf {\bibinfo {volume} {116}},\ \bibinfo {pages}
  {235302} (\bibinfo {year} {2016})}\BibitemShut {NoStop}%
\bibitem [{\citenamefont {Dast}\ \emph
  {et~al.}(2016{\natexlab{a}})\citenamefont {Dast}, \citenamefont {Haag},
  \citenamefont {Cartarius},\ and\ \citenamefont {Wunner}}]{Dast16a}%
  \BibitemOpen
  \bibfield  {author} {\bibinfo {author} {\bibfnamefont {D.}~\bibnamefont
  {Dast}}, \bibinfo {author} {\bibfnamefont {D.}~\bibnamefont {Haag}}, \bibinfo
  {author} {\bibfnamefont {H.}~\bibnamefont {Cartarius}}, \ and\ \bibinfo
  {author} {\bibfnamefont {G.}~\bibnamefont {Wunner}},\ }\href {\doibase
  10.1103/PhysRevA.93.033617} {\bibfield  {journal} {\bibinfo  {journal} {Phys.
  Rev. A}\ }\textbf {\bibinfo {volume} {93}},\ \bibinfo {pages} {033617}
  (\bibinfo {year} {2016}{\natexlab{a}})}\BibitemShut {NoStop}%
\bibitem [{\citenamefont {Dast}\ \emph
  {et~al.}(2016{\natexlab{b}})\citenamefont {Dast}, \citenamefont {Haag},
  \citenamefont {Cartarius}, \citenamefont {Main},\ and\ \citenamefont
  {Wunner}}]{Dast16b}%
  \BibitemOpen
  \bibfield  {author} {\bibinfo {author} {\bibfnamefont {D.}~\bibnamefont
  {Dast}}, \bibinfo {author} {\bibfnamefont {D.}~\bibnamefont {Haag}}, \bibinfo
  {author} {\bibfnamefont {H.}~\bibnamefont {Cartarius}}, \bibinfo {author}
  {\bibfnamefont {J.}~\bibnamefont {Main}}, \ and\ \bibinfo {author}
  {\bibfnamefont {G.}~\bibnamefont {Wunner}},\ }\href {\doibase
  10.1103/PhysRevA.94.053601} {\bibfield  {journal} {\bibinfo  {journal} {Phys.
  Rev. A}\ }\textbf {\bibinfo {volume} {94}},\ \bibinfo {pages} {053601}
  (\bibinfo {year} {2016}{\natexlab{b}})}\BibitemShut {NoStop}%
\bibitem [{\citenamefont {Dast}\ \emph {et~al.}(2014)\citenamefont {Dast},
  \citenamefont {Haag}, \citenamefont {Cartarius},\ and\ \citenamefont
  {Wunner}}]{Dast14a}%
  \BibitemOpen
  \bibfield  {author} {\bibinfo {author} {\bibfnamefont {D.}~\bibnamefont
  {Dast}}, \bibinfo {author} {\bibfnamefont {D.}~\bibnamefont {Haag}}, \bibinfo
  {author} {\bibfnamefont {H.}~\bibnamefont {Cartarius}}, \ and\ \bibinfo
  {author} {\bibfnamefont {G.}~\bibnamefont {Wunner}},\ }\href {\doibase
  10.1103/PhysRevA.90.052120} {\bibfield  {journal} {\bibinfo  {journal} {Phys.
  Rev. A}\ }\textbf {\bibinfo {volume} {90}},\ \bibinfo {pages} {052120}
  (\bibinfo {year} {2014})}\BibitemShut {NoStop}%
\bibitem [{\citenamefont {Breuer}\ and\ \citenamefont
  {Petruccione}(2002)}]{Breuer02a}%
  \BibitemOpen
  \bibfield  {author} {\bibinfo {author} {\bibfnamefont {H.-P.}\ \bibnamefont
  {Breuer}}\ and\ \bibinfo {author} {\bibfnamefont {F.}~\bibnamefont
  {Petruccione}},\ }\href@noop {} {\emph {\bibinfo {title} {{The theory of open
  quantum systems}}}}\ (\bibinfo  {publisher} {Oxford University Press},\
  \bibinfo {address} {Oxford},\ \bibinfo {year} {2002})\BibitemShut {NoStop}%
\bibitem [{\citenamefont {Anglin}(1997)}]{Anglin97a}%
  \BibitemOpen
  \bibfield  {author} {\bibinfo {author} {\bibfnamefont {J.}~\bibnamefont
  {Anglin}},\ }\href {\doibase 10.1103/PhysRevLett.79.6} {\bibfield  {journal}
  {\bibinfo  {journal} {Phys. Rev. Lett.}\ }\textbf {\bibinfo {volume} {79}},\
  \bibinfo {pages} {6} (\bibinfo {year} {1997})}\BibitemShut {NoStop}%
\bibitem [{\citenamefont {Fisher}\ \emph {et~al.}(1989)\citenamefont {Fisher},
  \citenamefont {Weichman}, \citenamefont {Grinstein},\ and\ \citenamefont
  {Fisher}}]{Fisher89a}%
  \BibitemOpen
  \bibfield  {author} {\bibinfo {author} {\bibfnamefont {M.~P.~A.}\
  \bibnamefont {Fisher}}, \bibinfo {author} {\bibfnamefont {P.~B.}\
  \bibnamefont {Weichman}}, \bibinfo {author} {\bibfnamefont {G.}~\bibnamefont
  {Grinstein}}, \ and\ \bibinfo {author} {\bibfnamefont {D.~S.}\ \bibnamefont
  {Fisher}},\ }\href {\doibase 10.1103/PhysRevB.40.546} {\bibfield  {journal}
  {\bibinfo  {journal} {Phys. Rev. B}\ }\textbf {\bibinfo {volume} {40}},\
  \bibinfo {pages} {546} (\bibinfo {year} {1989})}\BibitemShut {NoStop}%
\bibitem [{\citenamefont {Jaksch}\ \emph {et~al.}(1998)\citenamefont {Jaksch},
  \citenamefont {Bruder}, \citenamefont {Cirac}, \citenamefont {Gardiner},\
  and\ \citenamefont {Zoller}}]{Jaksch98a}%
  \BibitemOpen
  \bibfield  {author} {\bibinfo {author} {\bibfnamefont {D.}~\bibnamefont
  {Jaksch}}, \bibinfo {author} {\bibfnamefont {C.}~\bibnamefont {Bruder}},
  \bibinfo {author} {\bibfnamefont {J.~I.}\ \bibnamefont {Cirac}}, \bibinfo
  {author} {\bibfnamefont {C.~W.}\ \bibnamefont {Gardiner}}, \ and\ \bibinfo
  {author} {\bibfnamefont {P.}~\bibnamefont {Zoller}},\ }\href {\doibase
  10.1103/PhysRevLett.81.3108} {\bibfield  {journal} {\bibinfo  {journal}
  {Phys. Rev. Lett.}\ }\textbf {\bibinfo {volume} {81}},\ \bibinfo {pages}
  {3108} (\bibinfo {year} {1998})}\BibitemShut {NoStop}%
\bibitem [{\citenamefont {Anglin}\ and\ \citenamefont
  {Vardi}(2001)}]{Anglin01a}%
  \BibitemOpen
  \bibfield  {author} {\bibinfo {author} {\bibfnamefont {J.~R.}\ \bibnamefont
  {Anglin}}\ and\ \bibinfo {author} {\bibfnamefont {A.}~\bibnamefont {Vardi}},\
  }\href {\doibase 10.1103/PhysRevA.64.013605} {\bibfield  {journal} {\bibinfo
  {journal} {Phys. Rev. A}\ }\textbf {\bibinfo {volume} {64}},\ \bibinfo
  {pages} {013605} (\bibinfo {year} {2001})}\BibitemShut {NoStop}%
\bibitem [{\citenamefont {Vardi}\ and\ \citenamefont
  {Anglin}(2001)}]{Vardi01a}%
  \BibitemOpen
  \bibfield  {author} {\bibinfo {author} {\bibfnamefont {A.}~\bibnamefont
  {Vardi}}\ and\ \bibinfo {author} {\bibfnamefont {J.~R.}\ \bibnamefont
  {Anglin}},\ }\href {\doibase 10.1103/PhysRevLett.86.568} {\bibfield
  {journal} {\bibinfo  {journal} {Phys. Rev. Lett.}\ }\textbf {\bibinfo
  {volume} {86}},\ \bibinfo {pages} {568} (\bibinfo {year} {2001})}\BibitemShut
  {NoStop}%
\bibitem [{\citenamefont {Penrose}\ and\ \citenamefont
  {Onsager}(1956)}]{Penrose56a}%
  \BibitemOpen
  \bibfield  {author} {\bibinfo {author} {\bibfnamefont {O.}~\bibnamefont
  {Penrose}}\ and\ \bibinfo {author} {\bibfnamefont {L.}~\bibnamefont
  {Onsager}},\ }\href {\doibase 10.1103/PhysRev.104.576} {\bibfield  {journal}
  {\bibinfo  {journal} {Phys. Rev.}\ }\textbf {\bibinfo {volume} {104}},\
  \bibinfo {pages} {576} (\bibinfo {year} {1956})}\BibitemShut {NoStop}%
\bibitem [{\citenamefont {Yang}(1962)}]{Yang62a}%
  \BibitemOpen
  \bibfield  {author} {\bibinfo {author} {\bibfnamefont {C.~N.}\ \bibnamefont
  {Yang}},\ }\href {\doibase 10.1103/RevModPhys.34.694} {\bibfield  {journal}
  {\bibinfo  {journal} {Rev. Mod. Phys.}\ }\textbf {\bibinfo {volume} {34}},\
  \bibinfo {pages} {694} (\bibinfo {year} {1962})}\BibitemShut {NoStop}%
\bibitem [{\citenamefont {Cederbaum}\ and\ \citenamefont
  {Streltsov}(2003)}]{Cederbaum03a}%
  \BibitemOpen
  \bibfield  {author} {\bibinfo {author} {\bibfnamefont {L.}~\bibnamefont
  {Cederbaum}}\ and\ \bibinfo {author} {\bibfnamefont {A.}~\bibnamefont
  {Streltsov}},\ }\href {\doibase 10.1016/j.physleta.2003.09.058} {\bibfield
  {journal} {\bibinfo  {journal} {Phys. Lett. A}\ }\textbf {\bibinfo {volume}
  {318}},\ \bibinfo {pages} {564} (\bibinfo {year} {2003})}\BibitemShut
  {NoStop}%
\bibitem [{\citenamefont {Graefe}\ \emph {et~al.}(2008)\citenamefont {Graefe},
  \citenamefont {G{\"u}nther}, \citenamefont {Korsch},\ and\ \citenamefont
  {Niederle}}]{Graefe08b}%
  \BibitemOpen
  \bibfield  {author} {\bibinfo {author} {\bibfnamefont {E.~M.}\ \bibnamefont
  {Graefe}}, \bibinfo {author} {\bibfnamefont {U.}~\bibnamefont {G{\"u}nther}},
  \bibinfo {author} {\bibfnamefont {H.~J.}\ \bibnamefont {Korsch}}, \ and\
  \bibinfo {author} {\bibfnamefont {A.~E.}\ \bibnamefont {Niederle}},\ }\href
  {\doibase 10.1088/1751-8113/41/25/255206} {\bibfield  {journal} {\bibinfo
  {journal} {J. Phys. A}\ }\textbf {\bibinfo {volume} {41}},\ \bibinfo {pages}
  {255206} (\bibinfo {year} {2008})}\BibitemShut {NoStop}%
\bibitem [{\citenamefont {Baker}\ \emph {et~al.}(2005)\citenamefont {Baker},
  \citenamefont {Jessup},\ and\ \citenamefont {Manteuffel}}]{Baker05a}%
  \BibitemOpen
  \bibfield  {author} {\bibinfo {author} {\bibfnamefont {A.~H.}\ \bibnamefont
  {Baker}}, \bibinfo {author} {\bibfnamefont {E.~R.}\ \bibnamefont {Jessup}}, \
  and\ \bibinfo {author} {\bibfnamefont {T.}~\bibnamefont {Manteuffel}},\
  }\href {\doibase 10.1137/S0895479803422014} {\bibfield  {journal} {\bibinfo
  {journal} {SIAM J. Matrix Anal. Appl.}\ }\textbf {\bibinfo {volume} {26}},\
  \bibinfo {pages} {962} (\bibinfo {year} {2005})}\BibitemShut {NoStop}%
\bibitem [{\citenamefont {Walls}\ and\ \citenamefont
  {Milburn}(2008)}]{Walls08a}%
  \BibitemOpen
  \bibfield  {author} {\bibinfo {author} {\bibfnamefont {D.}~\bibnamefont
  {Walls}}\ and\ \bibinfo {author} {\bibfnamefont {G.~J.}\ \bibnamefont
  {Milburn}},\ }\href {\doibase 10.1007/978-3-540-28574-8} {\emph {\bibinfo
  {title} {{Quantum Optics}}}},\ \bibinfo {edition} {2nd}\ ed.\ (\bibinfo
  {publisher} {Springer},\ \bibinfo {address} {Berlin},\ \bibinfo {year}
  {2008})\BibitemShut {NoStop}%
\end{thebibliography}
\end{document}